\documentclass[aps,prl,twocolumn,prl,showpacs]{revtex4}
\usepackage[latin9]{inputenc}
\usepackage{color}
\usepackage[colorlinks,urlcolor=blue,citecolor=blue,linkcolor=blue]{hyperref}
\usepackage{verbatim}
\usepackage{float}
\usepackage{amsmath}
\usepackage{amssymb}
\usepackage{graphicx}
\usepackage{esint}
\begin{document}

\title{Observation of a Strong Atom-Dimer Attraction in a Mass-Imbalanced Fermi-Fermi Mixture}

\author{Michael Jag$^{1,2}$}
\author{Matteo Zaccanti$^{1,3}$}
\author{Marko Cetina$^{1}$}
\author{Rianne S. Lous$^{1,2}$}
\author{Florian Schreck$^{1}$}
\author{Rudolf Grimm$^{1,2}$}
\affiliation{$^1$Institut f\"ur Quantenoptik und Quanteninformation (IQOQI), \"Osterreichische Akademie der Wissenschaften}
\affiliation{$^2$Institut f\"ur Experimentalphysik, Universit\"at Innsbruck, 6020 Innsbruck, Austria}
\affiliation{$^3$CNR Istituto Nazionale Ottica, 50019 Sesto Fiorentino, Italy}
\author{Dmitry S. Petrov$^{1}$}
\author{Jesper Levinsen$^{2,3}$}
\affiliation{$^1$Universit\a'e Paris-Sud, CNRS, LPTMS, UMR8626, Orsay, F-91405, France}
\affiliation{$^2$TCM group, Cavendish Laboratory, JJ Thomson Avenue, Cambridge, CB3 0HE, United Kingdom}
\affiliation{$^3$Aarhus Institute of Advanced Studies, Aarhus University, DK-8000 Aarhus C, Denmark}

\date{\today}


\begin{abstract}
We investigate a mixture of ultracold fermionic $^{40}$K atoms and weakly bound $^{6}$Li$^{40}$K dimers on the repulsive side of a heteronuclear atomic Feshbach resonance. 
By radio-frequency spectroscopy we demonstrate that the normally repulsive atom-dimer interaction is turned into a strong attraction.
The phenomenon can be understood as a three-body effect in which two heavy $^{40}$K fermions exchange the light $^{6}$Li atom, leading to attraction in odd partial-wave channels (mainly $p$-wave). 
Our observations show that mass imbalance in a fermionic system can profoundly change the character of interactions as compared to the well-established mass-balanced case.
\end{abstract}

\maketitle

Ultracold fermions with tunable interactions provide remarkable possibilities to model the many-body physics of strongly interacting states of quantum matter under well-controllable conditions \cite{Giorgini2008tou, Bloch2008mbp}.
Fermionic superfluids, realized by combining two different spin states of a fermionic atomic species and controlling their $s$-wave interaction through a Feshbach resonance \cite{Chin2010fri}, have led to spectacular achievements. 
Beyond these experimentally well-established fermionic systems, mass imbalance offers an additional degree of freedom, with interesting prospects for new many-body phenomena having no counterpart in the mass-balanced case, such as novel quantum phases or superfluid states in various trapping environments \cite{Iskin2006tsf,Bausmerth2009ccl,Gezerlis2009hlf,Keyserlingk2011ifi,Sotnikov2012aom,Cui2013psi,Gubbels2009lpi, Mathy2011tma, Qi2012hpf, Daily2012tot, Petrov2007cpo, Baranov2008spb, Sanchez1991tcf, Orso2010llo, Dalmonte2012dta, Nishida2008ufg, Nishida2009cie}.

Striking effects of mass imbalance in fermionic systems already emerge at the few-body level. 
A resonantly interacting three-body system of one light ($\downarrow$) and two heavy ($\uparrow$) fermions is known to exhibit bound states depending on the mass ratio $m_{\uparrow}/m_{\downarrow}$. 
While Efimov trimer states require large mass ratios ($m_{\uparrow}/m_{\downarrow} > 13.6$), for repulsive interactions, non-Efimovian trimer states can exist in an intermediate regime ($13.6 > m_{\uparrow}/m_{\downarrow} > 8.17$) \cite{Kartavtsev2007let}. 
Below the critical value of $8.17$, the last state turns into an atom-dimer scattering resonance in the $p$-wave channel \cite{Kartavtsev2007let}.

The $^{40}$K-$^6$Li mixture serves as the prime system for current experiments on tunable mass-imbalanced Fermi-Fermi mixtures \cite{Wille2008eau, Costa2010swi, Trenkwalder2011heo}. 
The corresponding mass ratio of $m_{\uparrow}/m_{\downarrow} \approx 6.64$ lies well in the regime of near-resonant atom-dimer interactions \cite{Levinsen2009ads, Levinsen2011ada}:
as the most prominent effect, theory predicts a substantial attraction resulting from higher partial waves (mainly $p$-wave) in a regime where one would naively, based on $s$ waves alone, expect a strong repulsion. 
This also makes the corresponding many-body problem in a $^{40}$K-$^6$Li mixture significantly more complicated and much richer than in the widely investigated mass-balanced case.

In this Letter, we investigate the interaction between $^{40}$K atoms and weakly bound $^6$Li$^{40}$K dimers near an interspecies Feshbach resonance (FR). 
We employ radio-frequency (rf) spectroscopy by using two different internal states of $^{40}$K, one strongly interacting with the dimers and the other one practically non-interacting \cite{Kohstall2012mac}. 
We observe line shifts and collisional broadening and interpret these in terms of the real and imaginary part of the forward-scattering amplitude $f(0)$ for atom-dimer collisions, calculated on the basis of the theoretical approach of Ref.~\cite{Levinsen2011ada}. 
The comparison between theory and experiment shows excellent agreement and, in particular, demonstrates the predicted atom-dimer attraction on the repulsive side of the interspecies FR.

\begin{figure}
\vskip 1.2pt \includegraphics[clip,width=0.98\columnwidth]{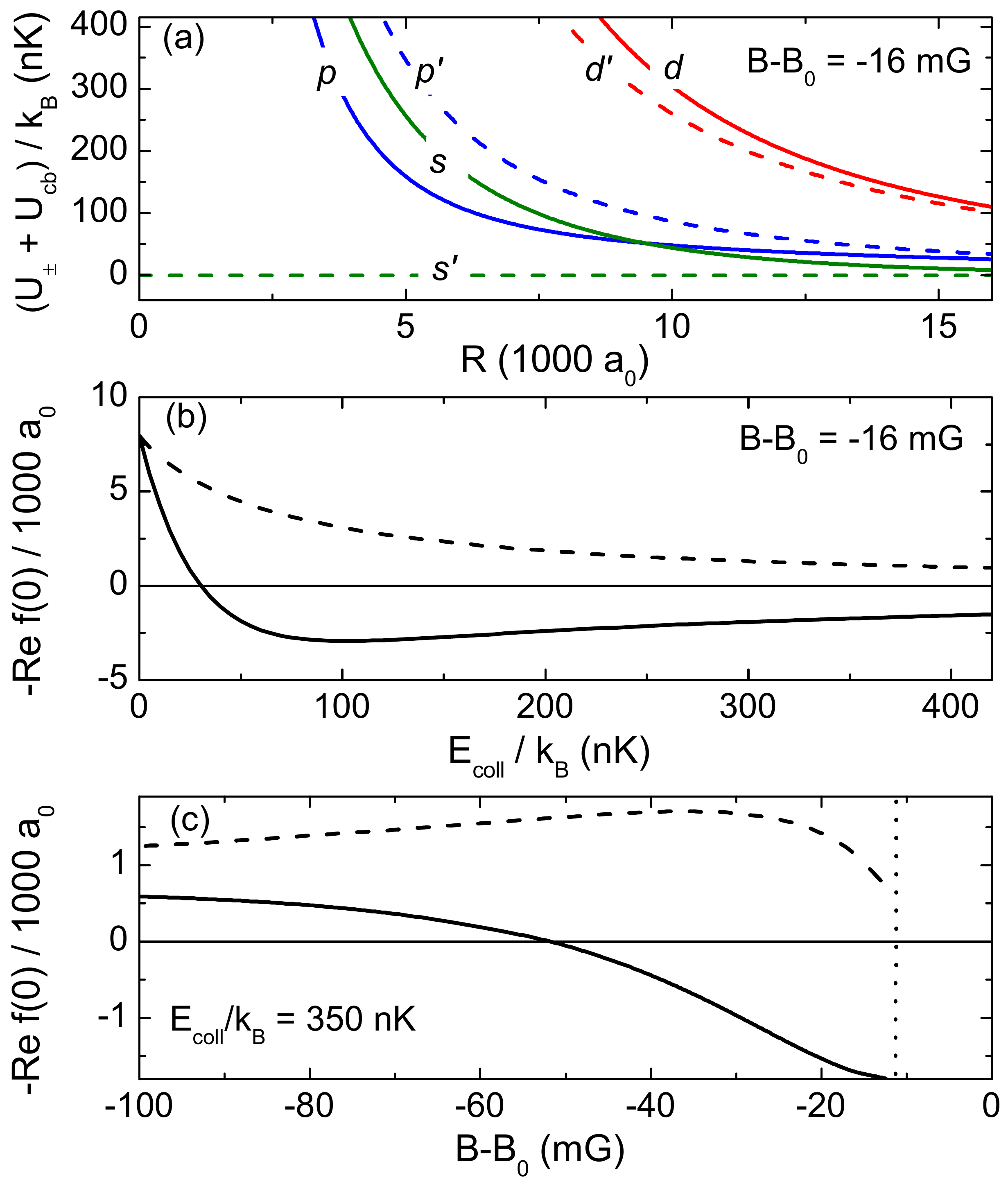}
\caption{Interaction between $^{40}$K atoms and $^{6}$Li$^{40}$K dimers near the $155\,$G interspecies FR. 
(a) Total interaction potentials as a function of the distance $R$ between the two K atoms for the $s$, $p$, and $d$ channels (dashed curves with labels $s^{\prime}$, $p^{\prime}$, $d^{\prime}$ refer to the unmodified centrifugal barriers).
Here we have chosen a magnetic detuning of $B-B_0=-16\,$mG, corresponding to a $s$-wave scatering length of $a = 3528\,a_0$ and to a dimer binding energy of $E_{\rm b} / k_{\rm B} = 600\,$nK.
(b) Real part of the forward-scattering amplitude $f(0)$ as a function of the collision energy $E_{\rm coll}$ (solid line) in comparison with the $s$-wave contribution (dashed line).
(c) Same as in (b), but as a function of the magnetic detuning $B-B_0$ for a fixed collision energy $E_{\rm coll} / k_{\rm B} = 350\,$nK.
The dotted line indicates the dimer breakup threshold, $E_{\rm coll} = E_{\rm b}$.}
\label{Figure1}
\end{figure}

The interaction of a heavy atom with a heavy-light dimer can be understood in the Born-Oppenheimer approximation, where the atom-dimer potentials are taken to be the eigenenergies of the light atom for a given separation $R$ between the heavy ones. 
As in the usual double-well problem with tunneling, the state localized near one heavy atom is mixed with the state localized near the other; 
the symmetric and antisymmetric superpositions lead to the attractive $U_+(R)<0$ and repulsive $U_-(R)>0$ potentials, respectively. 
Note the analogy to the well-known H$_2^+$ cation, where the exchange of the electron leads to a symmetric bound state and an antisymmetric unbound state \cite{pauling1928tao}. 
In our experiment, the heavy particles are identical fermions, making the atom-dimer interaction channel dependent.
The symmetric (antisymmetric) state corresponds to odd (even) values of the total angular momentum $l$ \cite{Levinsen2011ada}. 
In Fig.~\ref{Figure1}(a) we plot the total effective potentials $U_\pm+U_{\rm cb}$ (solid lines) and the bare centrifugal barriers $U_{\rm cb}=l(l+1)\hbar^{2}/m_{\uparrow}R^{2}$ (dashed lines) for $l = 0$, 1, and 2 (i.e., $s$-, $p$-, and $d$-wave channels) for typical experimental conditions. 
At distances on the order of typical de Broglie wavelength, $U_\pm$ can be comparable to $U_{\rm cb}$ and we expect significant interaction effects in non-zero partial waves.

The relevant quantity that characterizes the net effect of all partial waves is the atom-dimer forward scattering amplitude \cite{sobelman1973ait, baranger1958sqm, baranger1958git},
\begin{equation}
\label{f0}
f(0)=\sum_{l=0}^{\infty}(2l+1)\left[\frac{\sin2\delta_l(k_{\rm coll})}{2k_{\rm coll}}+i\frac{\sin^2\delta_l(k_{\rm coll})}{k_{\rm coll}}\right],
\end{equation} 
where $k_{\rm coll}=\sqrt{2\mu_3 E_{\rm coll}}/\hbar$ is the wavenumber associated with the relative atom-dimer motion and $\mu_3$ is the reduced atom-dimer mass. 
The phase shifts $\delta_l$ for the three lowest partial waves have been computed in Ref.~\cite{Levinsen2011ada} and here we extend the result to higher ones since they give significant contributions \cite{MJSupMat}.
In Fig.~\ref{Figure1}(b) we show the resulting $-{\rm Re}\, f(0)$ as a function of the collision energy $E_{\rm coll}$ for the same conditions as in Fig.~\ref{Figure1}(a). 
In the limit of $E_{\rm coll} \rightarrow 0$, the quantity $-{\rm Re}\,f(0)$ corresponds to the atom-dimer $s$-wave scattering length.
At $E_{\rm coll} \ll 0.1E_{\rm b}$, with $E_{\rm b}$ being the dimer binding energy, $s$-wave scattering (dashed line) dominates and the net interaction is repulsive, $-{\rm Re}\, f(0)>0$.

For $E_{\rm coll} \gtrsim 0.1E_{\rm b}$, higher partial-wave contributions lead to a sign reversal of ${\rm Re}\, f(0)$, changing the character of the interaction from repulsive into attractive. 
This sign reversal also appears if, at a fixed collision energy, the magnetic detuning from the FR center is varied, see Fig.~\ref{Figure1}(c).
In the realistic example of Fig.~\ref{Figure1}(c) the sign reversal takes place at a magnetic detuning of $B - B_0 = -53\,$mG, where the binding energy is $E_{\rm b}/k_{\rm B} \approx 3.1\,\mu$K, corresponding to roughly ten times the collision energy $E_{\rm coll}/k_{\rm B} = 350\,$nK.
The theory lines in Fig.~\ref{Figure1}(c) stop close to the FR center at the magnetic field detuning where $|E_{\rm b}| = E_{\rm coll}$ (dotted line), beyond which the inelastic channel of collisional dimer dissociation opens up.

The starting point of our experiments is an optically trapped, near-degenerate Fermi-Fermi mixture of typically $4 \times 10^4$ $^{40}$K atoms and $1 \times 10^5$ $^6$Li atoms. 
The preparation procedures are described in our previous work \cite{Spiegelhalder2010aop, Trenkwalder2011heo}. 
We choose a particular FR that occurs between Li atoms in the lowest Zeeman sub-level Li$|1\rangle$ ($f=1/2, m_f=+1/2$) and K atoms in the third-to-lowest sub-level K$|3\rangle$ ($f=9/2, m_f=-5/2$) \cite{Naik2011fri}. 
The $s$-wave interspecies scattering length $a$ can be magnetically tuned as $a = a_{\rm bg}[1 - \Delta/(B-B_0)]$ with $a_{\rm bg} = 63.0\,a_0$ ($a_0$ is Bohr's radius) and $\Delta = 880\,$mG \cite{Naik2011fri}. 
The resonance is rather narrow, as characterized by the length parameter $R^{*}=2700\,a_0$ \cite{Petrov2004tbp}.
The position of the FR center near $B \approx 154.7\,$G depends on the trap setting, as it includes small shifts induced by the trapping light. 
For each trap setting we have calibrated the FR center $B_0$ with $\leq 2\,$mG accuracy \cite{MJSupMat}. 

We create an atom-dimer mixture by a Feshbach ramp across the resonance and by subsequent purification and spin-manipulation techniques \cite{MJSupMat}. 
While the dimers are formed in the Li$|1\rangle$-K$|3\rangle$ spin channel, we initially prepare the free atoms in the second-to-lowest spin state K$|2\rangle$ ($f=9/2, m_f=-7/2$), for which the interaction with the dimers is negligible. 
The total number of dimers and atoms is $1.5 \times 10^4$ and $7 \times 10^3$, respectively. 
The interspecies attraction during the Feshbach ramp results in a collective oscillation of the dimer cloud, which we can take into account by introducing an effective temperature $T_{\rm eff}$ \cite{MJSupMat}. 
We use three different trap settings, for which $T_{\rm eff} = 165\,$nK, $232\,$nK, and $370\,$nK.
This corresponds to mean dimer densities as experienced by the atoms of $\bar{n}_{\rm D}=5.2\times 10^{11}\,$cm$^{-3}$, $8.2\times 10^{11}\,$cm$^{-3}$, and $1.4\times 10^{12}\,$cm$^{-3}$, respectively.

To investigate the interaction between the K$|3\rangle$ atoms and the Li$|1\rangle$K$|3\rangle$ dimers, we carry out rf spectroscopy. 
This can be done in two different ways, either by driving the K atoms from the noninteracting state $|2\rangle$ into the interacting state $|3\rangle$ (method A) or vice versa (method B). 
With our K atoms initially prepared in the state $|2\rangle$, we carry out method A by applying a 1-ms rf pulse. 
For method B, we rapidly transfer the full K$|2\rangle$ population into K$|3\rangle$ using a short 90-$\mu$s preparation pulse without spectral resolution, and then drive the spectrally resolving transition with a 1-ms pulse. 
Our signal in both cases is the fraction of transferred atoms as a function of the rf detuning $\nu-\nu_0$ with respect to the unperturbed transition frequency $\nu_0$, the latter being determined by the rf spectroscopy in the absence of dimers. 

\begin{figure}
\vskip 0 pt \includegraphics[clip,width=0.98\columnwidth]{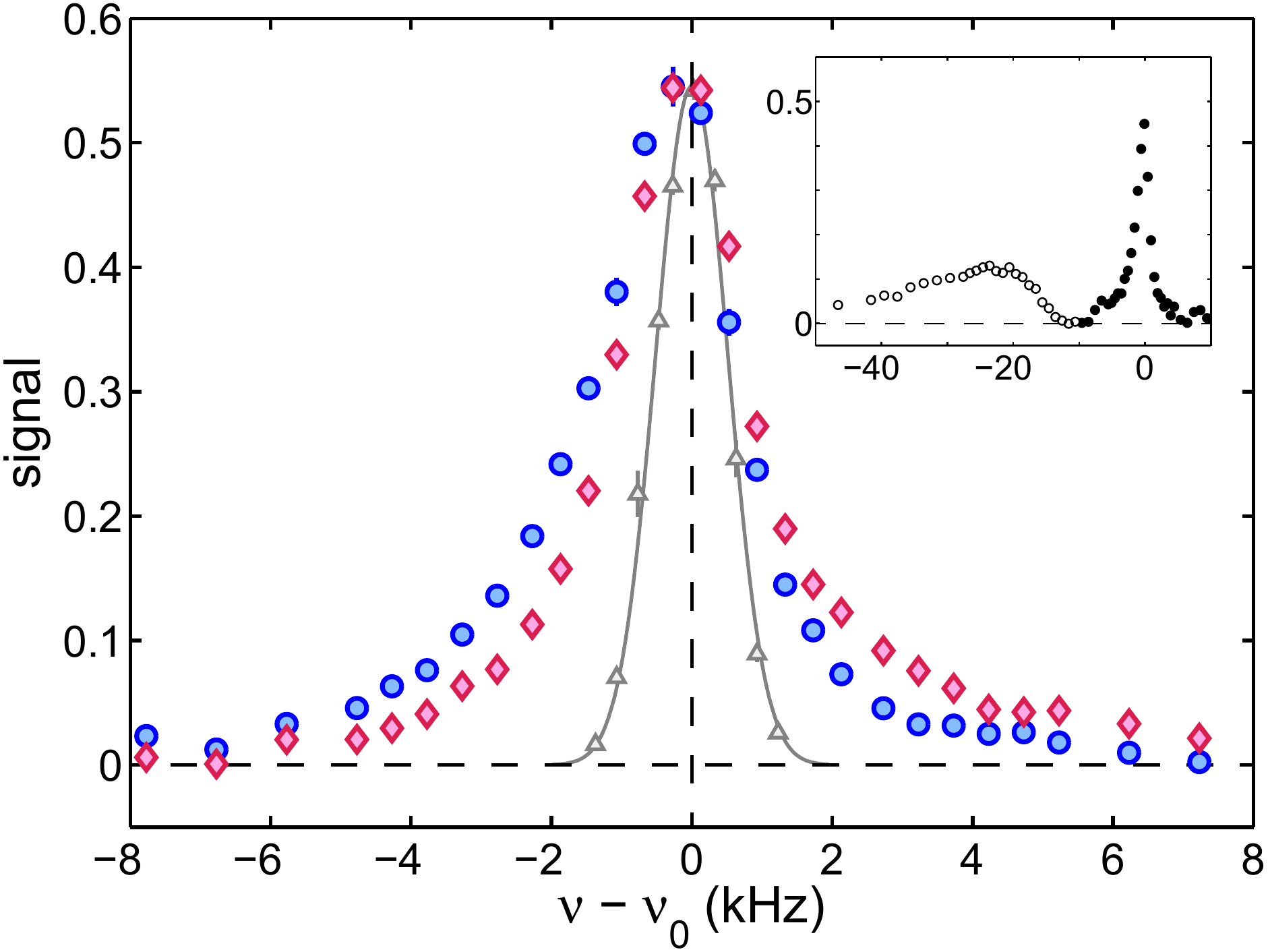}
\caption{Sample rf spectra taken at $B-B_0 = -20\,$mG at $T_{\rm eff}=232\,$nK. 
The red diamonds (blue circles) show data recorded using method A (B). 
For reference, the gray triangles show data recorded in the absence of dimers together with a Gaussian fit (gray line). 
Inset: Spectrum at $-17\,$mG over an extended frequency range. 
The molecular dissociation signal (open symbols), recorded with $30\times$ increased rf power, is clearly separated from the atomic peak (filled symbols).}
\label{Figure2} 
\end{figure}

Sample spectra, at a magnetic detuning of $B-B_0 = -20\,$mG, are shown in Fig.~\ref{Figure2}.
The spectra recorded by methods A and B (circles and diamonds in Fig.~\ref{Figure2}) show both a broadening and a peak shift, as compared to the spectra recorded in the absence of dimers (triangles). 
Although the spectra very close to the FR center reveal asymmetries in their wings, which depend on the method applied, their peak shifts and broadenings are consistent for both methods.
In the range of detunings $B-B_0$ studied in the present work the molecular dissociation signal is always well separated from the atomic line (inset of Fig.~\ref{Figure2}), and thus does not affect the lineshape of the atomic signal.

\begin{figure}
\vskip 0 pt \includegraphics[clip,width=1\columnwidth]{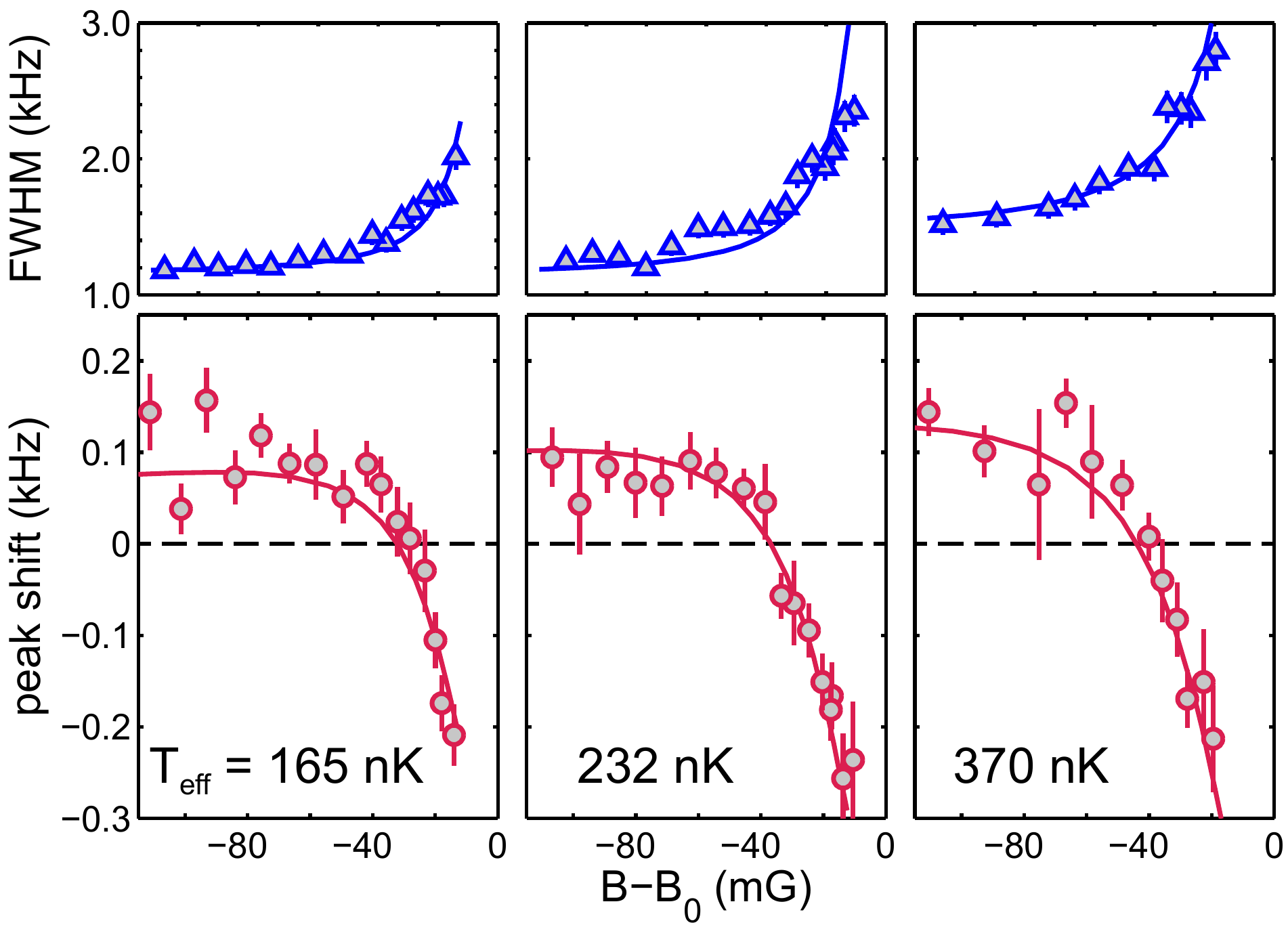}
\caption{Widths (blue triangles) and peak shifts (red circles) extracted from the rf spectra as a function of the magnetic field detuning $B-B_{\rm 0}$ for the three different values of $T_{\rm eff}$. The lines are the corresponding theoretical predictions. 
To account for fluctuations  in the dimer number of different spectra, the widths and peak shifts are scaled to a dimer number of $15,000$, which is typical for all spectra.}
\label{Figure3} 
\end{figure}

Figure~\ref{Figure3} shows the widths and peak shifts \cite{MJEndnote} of the rf spectroscopic signal, recorded by method A, as a function of $B -B_0$ for our three values of $T_{\rm eff}$. 
When the FR center is approached, the spectrum broadens and its peak shifts from a positive to a negative rf detuning. 
With increasing temperature, the corresponding zero crossing shows a trend to move towards larger detunings.

We interpret the obtained results in the framework of the impact theory of pressure-induced effects on spectral lines, which assumes the collisions to be effectively instantaneous.
This theory predicts Lorentzian profiles centered near the unperturbed frequency $\nu_0$ whose line shifts and broadenings are proportional to the real and imaginary parts of the thermally averaged atom-dimer forward scattering amplitude $f(0)$ \cite{sobelman1973ait, baranger1958sqm, baranger1958git}, respectively. 
The real part of $f(0)$ shifts the energy of the K atoms, causing an average shift in the frequency of their peak rf response of $\delta\nu=-\hbar\bar{n}_{\rm D}{\rm Re}\langle f(0)\rangle/\mu_{3}$, where $\langle f(0)\rangle$ denotes the thermal average of $\, f(0)$ over all atom-dimer collision energies $E_{\rm coll}$. 
The red solid lines in Fig.~\ref{Figure3} show the theoretical results for $\delta\nu$ for the respective molecule densities and collision energies. 
The optical theorem relates the imaginary part of $f(0)$ to the average elastic scattering rate $\tau^{-1}$ as $\tau^{-1}=4\pi\hbar\bar{n}_{\rm D}{\rm Im}\langle f(0)\rangle /\mu_3$. 
The resulting finite lifetime $\tau$ of the atoms' wavepackets causes Lorentzian broadening with a full-width at half-maximum (FWHM) $1/2\pi\tau$. 
The blue solid lines in Fig.~\ref{Figure3} show the predicted FWHM, including additional broadening due to the finite duration of our rf pulse \cite{MJEndnote2}. 

The collisional broadening yields information on the elastic scattering rate. 
At typical detunings of $B - B_0 \approx -20\,$mG, our data show an elastic atom-dimer scattering rate on the order of $1/(100\,\mu$s). 
A comparison with the observed dimer decay rate of about $1/(5\,$ms) gives a lower limit for the ratio of elastic to inelastic atom-dimer collisions of 50.
We note that in our system the dimers spontaneously dissociate on a timescale of about $10\,$ms \cite{Naik2011fri}.

The comparison between the experimentally observed and the theoretically calculated line shifts and broadenings shows remarkable agreement over the whole parameter range investigated. 
The somewhat asymmetric spectral wings are beyond the impact theory \cite{szudy1996pol} and thus cannot be reproduced. 
Indeed, a substantial contribution to the wings comes from the photon emission/absorption events for which K atoms find themselves inside the atom-dimer interaction range, i.e.\ {\it during atom-dimer collisions}, which are assumed instantaneous in the impact theory. 
It is then understood that, for example, the left ``attractive'' wing of the B-spectrum is larger than that of the A-spectrum.
Since in the former case potassium atoms are initially attracted by dimers, the probability to find them near dimers is enhanced. 
Effects that are beyond the impact theory become more pronounced as we approach the FR because of the increased atom-dimer collision time.

\begin{figure}
\vskip 0 pt \includegraphics[clip,width=1\columnwidth]{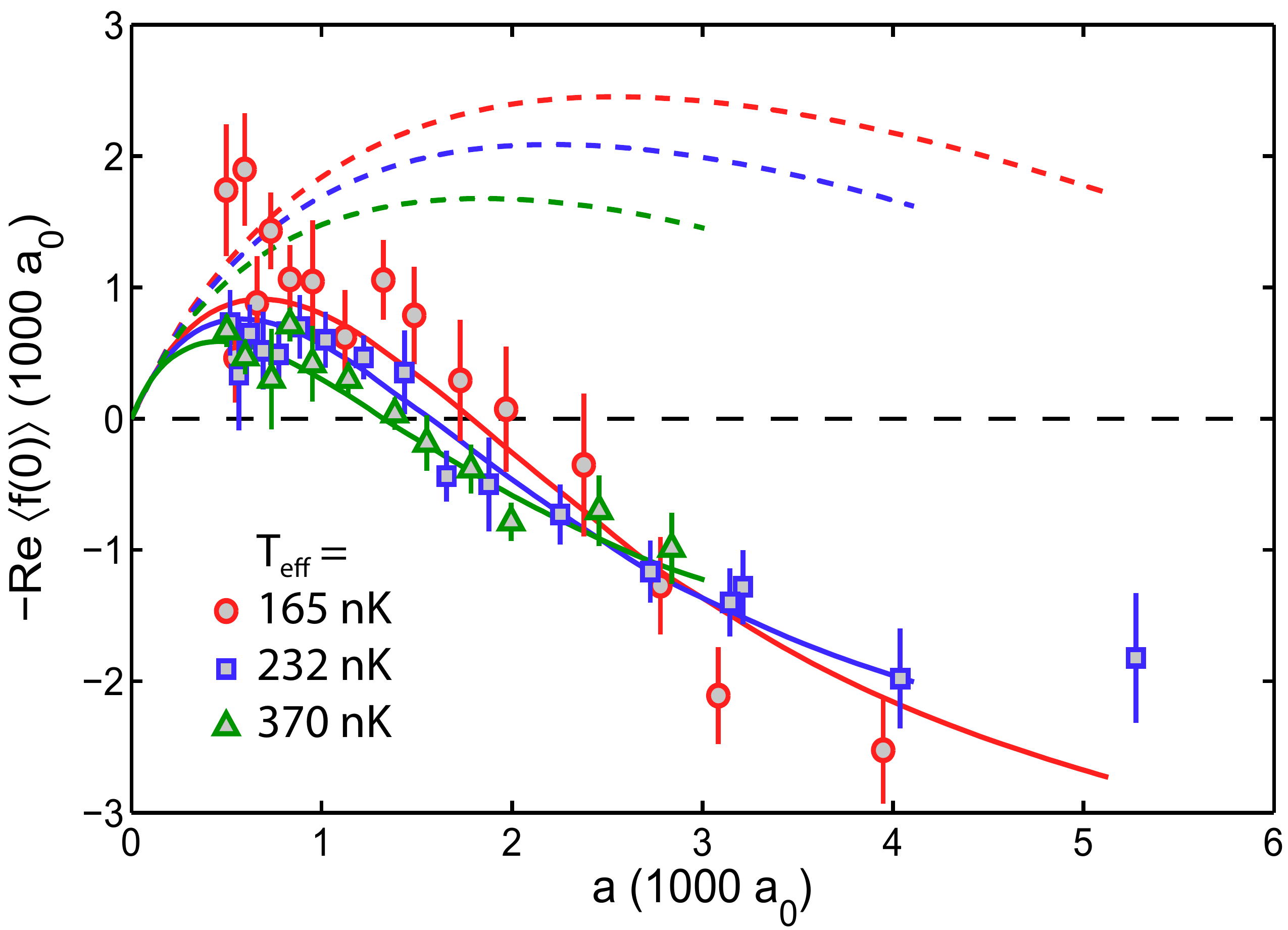}
\caption{Real part of the atom-dimer forward-scattering amplitude as a function of the atom-atom scattering length $a$ for the three different values of $T_{\rm eff}$.
The symbols and the lines show the data and the theoretical predictions from Fig.~\ref{Figure3}. 
For comparison, the dashed lines indicate the respective $s$-wave contributions. 
The theoretical lines stop at $k_{\rm B}T_{\rm eff} = E_{\rm b}$/2.}
\label{Figure4} 
\end{figure}

We finally discuss the interaction strength in our mixture in terms of $-{\rm Re}\langle f(0)\rangle$, which characterizes the interactions in a way that is analogous to $a$ in the $s$-wave mean-field picture. 
We use the experimental peak-shift data from Fig.~\ref{Figure3} to extract $-{\rm Re}\langle f(0)\rangle$ and plot it together with the corresponding theoretical results in Fig.~\ref{Figure4}. 
The sign reversal shows up for values of $a$ being somewhat below $2000\,a_0$, with the expected temperature dependence of the zero crossing. 
For $a \approx 4000\,a_0$, the attractive interaction already corresponds to about $-2000\,a_0$. 
For even larger values of $a$, we would enter the more complicated regime of collisional dimer dissociation, which is beyond the scope of the present investigations. 
We note, however, that rf spectra acquired more deeply in the strongly interacting regime show strongly asymmetric lineshapes and have peaks shifted to even larger negative detunings.

In conclusion, we have demonstrated a three-body phenomenon in a mixture of heavy and light fermions, which leads to a sign reversal of the atom-dimer interaction near a FR, turning repulsion into a strong attraction. 
The effect is due to higher partial-wave (mainly $p$-wave) contributions, which are present even at very low collision energies in the nanokelvin regime. 
Remarkably, this few-body effect changes the character of the interaction without introducing detrimental losses. 
In contrast to few-body phenomena of the Efimov type \cite{Ferlaino2011eri}, the centrifugal barrier still protects the atoms from approaching each other too closely. 
The resulting collisional stability is a promising feature for many-body physics in Fermi-Fermi mixtures.

Our work lays the ground for a wealth of future studies on mass-imbalanced fermionic mixtures in the strongly interacting regime. 
Asymmetric phases with coexisting dimers and heavy atoms are energetically favored in a way not present in mass-balanced systems \cite{Qi2012hpf}. 
Related mechanisms in quantum-degenerate situations may lead to exotic new many-body effects, including the emergence of imbalanced superfluids \cite{Qi2012hpf}, the condensation into non-zero momentum states \cite{Mathy2011tma}, and the appearance of $p$-wave superfluidity of heavy atoms mediated by light atoms \cite{Nishida2009cia}. 
On the few-body side, a direct prospect for our K-Li system is to confine the K atoms in an optical lattice, which is predicted to lead to the formation of stable trimer states \cite{Petrov2007cpo, Nishida2009cie, Levinsen2009ads}.

We acknowledge funding by the Austrian Science Fund FWF with SFB FoQuS (F40-P04). 
M.Z. was supported by the FWF within the Lise Meitner program (M1318), 
D.S.P. by the Institut Francilien de Recherche sur les Atomes Froids (IFRAF), 
and J.L. acknowledges support from the Carlsberg Foundation.

\vspace{-0.1cm}
\section*{Supplemental Material}
\vspace{-0.3cm}
\subsection{Light shift of the Feshbach resonance}
\label{lightshift}

The Feshbach resonance (FR) that we employ for tuning the interactions in our system occurs between $^{6}$Li atoms in their lowest internal state, denoted Li$|1\rangle$ ($f = 1/2, m_f = +1/2$), and $^{40}$K atoms in their third-to-lowest state K$|3\rangle$ ($f = 9/2, m_f = -5/2$). 
This resonance has been investigated in detail in Ref.~\cite{Naik2011fri}. 
The magnetic field dependent Li-K $s$-wave scattering length is given by
\begin{equation}
    a(B)= a_{\rm bg} \left(1- \frac{\Delta}{B-B_0}\right)
\end{equation}
where $a_{\rm bg} = 63.0\,a_0$ is the background scattering length, $\Delta=0.88\,$G is the width, and $B_0$ is the center of the resonance near $154.7\,$G. 

As we already pointed out in Ref.~\cite{Kohstall2012mac}, the optical trap induces a differential light shift between the atom pair state and the molecular state.
This leads to a light-induced shift of the FR center.
For the experiments presented in the main text, we use a near-infrared laser with a wavelength of $1064\,$nm (single-mode operation) in three different trap settings.
Therefore, the center of the FR needs to be determined for each trap setting.

To determine $B_0$ we perform radio-frequency (rf) spectroscopy of the Feshbach molecules. 
For each trap setting, this is done in the following way: 
We prepare a nonresonant mixture of Li atoms in state Li$|1\rangle$ and K atoms in their second-to-lowest state K$|2\rangle$ several tens of mG below the approximate position of the resonance center. 
Here, we apply a strong 500-$\mu$s rf pulse at a variable frequency $\nu$, several kHz below the unperturbed K$|2\rangle$$\rightarrow $K$|3\rangle$ transition frequency $\nu_0$. 
This pulse drives Li$|1\rangle$-K$|2\rangle$ atom pairs into the Li$|1\rangle$K$|3\rangle$ dimer state.
To determine the number of dimers associated, we subsequently dissociate the dimers into a Li$|1\rangle$ and a K$|3\rangle$ atom by a 300-$\mu$s magnetic field ramp to $154.8\,$G. 
By recording absorption images we then determine the populations $N_2$ and $N_3$ of the K spin states K$|2\rangle$ and K$|3\rangle$, respectively.

\begin{center}
\begin{figure}
\includegraphics[width=1 \columnwidth, clip]{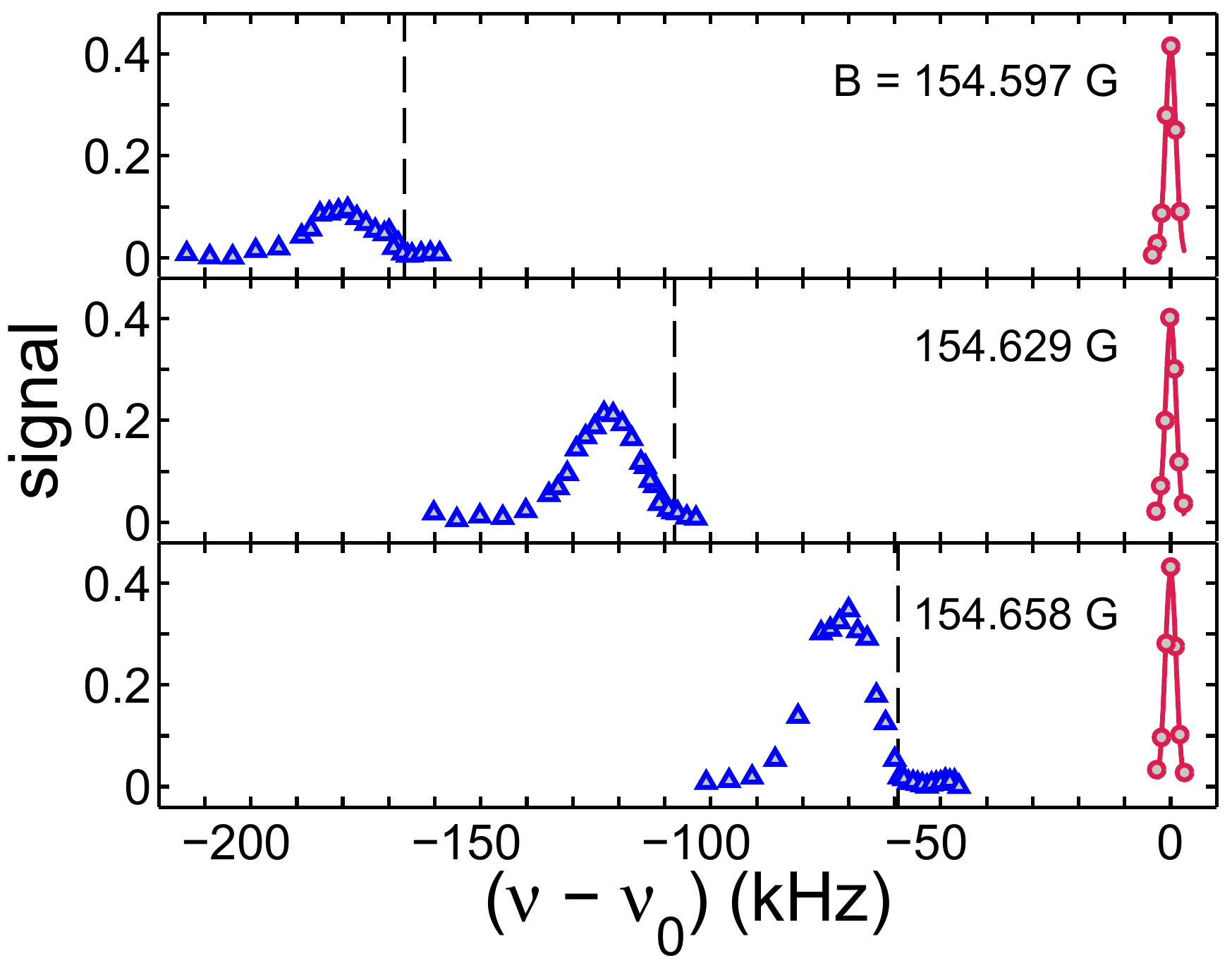}
\caption{Data from the molecular rf association spectroscopy in trap 2. 
Red circles were taken with a rf power set to the value to match the $\pi$-pulse condition in the absence of interactions (no Li$|1\rangle$ present) and is scaled by $0.5$.
Blue points were taken with a $30\times$ larger rf power. 
The dashed lines indicate the binding energy $E_{\rm b}(B)$.}
\label{FigureSI1}
\end{figure}
\end{center}

\begin{center}
\begin{figure}
\includegraphics[width=1 \columnwidth, clip]{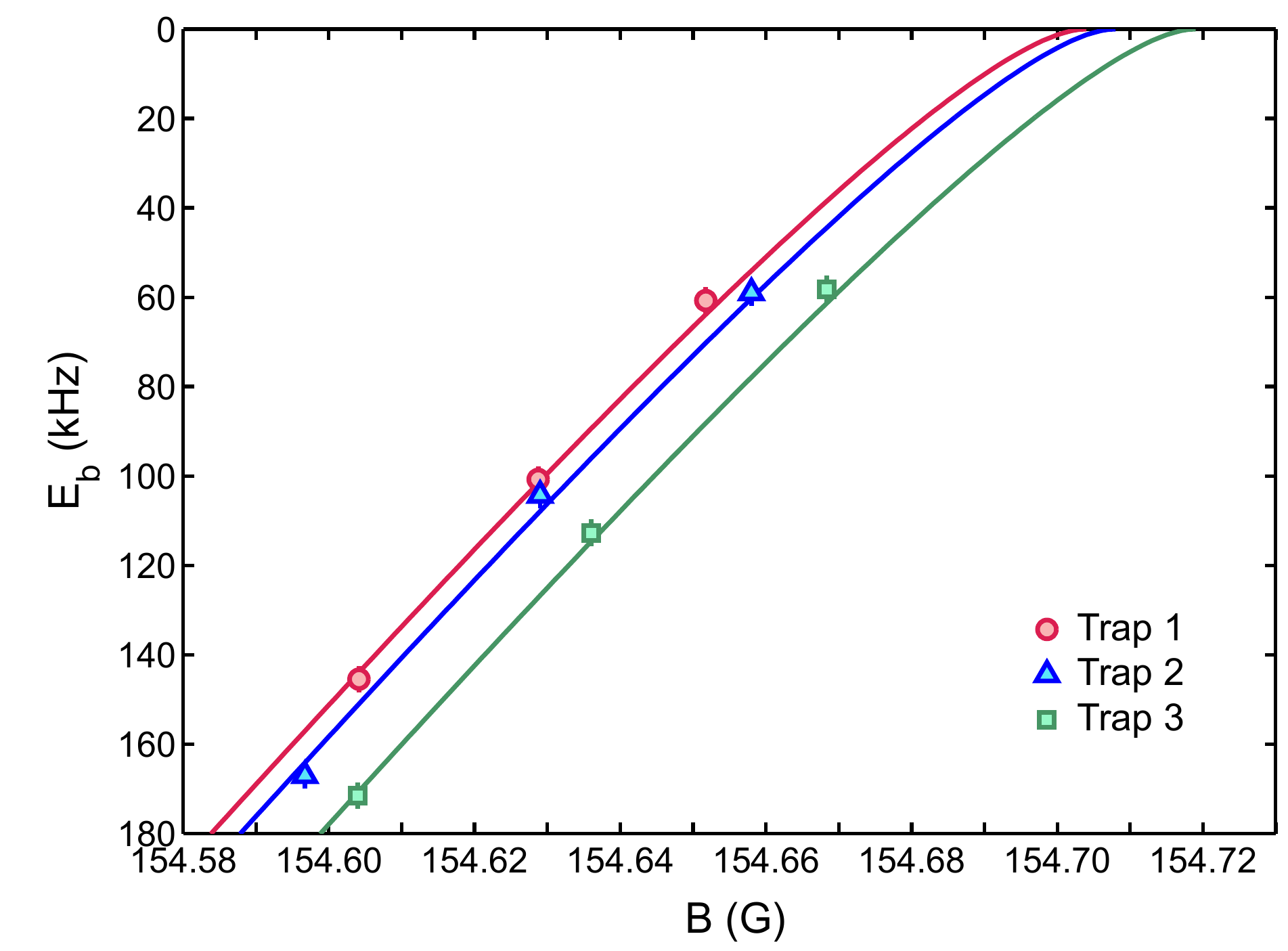}
\caption{Determination of the FR center $B_0$ by rf association of dimers. 
The points show the experimentally determined molecular binding energies $E_{\rm b}(B)$ for the three trap settings. 
The solid curves are fits of a theoretical model (see text) to the experimental data.}
\label{FigureSI2}
\end{figure}
\end{center}

\vspace{-1.7cm}
By plotting the signal, given by $N_3/(N_3+N_2)$, against the rf detuning $\nu - \nu_0$, we resolve the molecule association spectrum; see Fig.~\ref{FigureSI1}.
The unperturbed transition frequency $\nu_0$, corresponding to the Zeeman splitting of the two states, is determined by rf spectroscopy in the absence of Li$|1\rangle$ (red points).
We determine the binding energy of the molecules from the onset frequency of the molecular association spectra. 
As the onset frequency, we use the upper rf frequency at which the fraction of atoms transferred is roughly $10\%$ of its peak height.
We have checked that, within the errors of our measurements, this criterion agrees with the result obtained by fitting the line-shape model \cite{Chin2005rft} to the spectra, as was done in Ref.~\cite{Kohstall2012mac}.
This procedure is repeated for each trap power at various magnetic fields. 

We then fit a model \cite{Kohstall2012mac} for the molecular binding energy near our FR to the data with $B_0$ as the only free parameter; see Fig.~\ref{FigureSI2}. 
This procedure allows us to determine the resonance center in each trap setting with an uncertainty of $\pm 2\,$mG.
The accuracy of our determination of the resonance position is limited by the uncertainty in the FR parameters \cite{Naik2011fri} used in the model for the binding energy.
We determine the center of the FR of trap 1, 2, and 3 to be at the magnetic field of $154.704\,$G, $154.708\,$G, and $154.719\,$G, respectively.

\vspace{-0.4cm}
\subsection{Preparation of the atom-dimer mixture}
\label{preparation}

To cool our atomic sample, we evaporate a Li$|1\rangle$-Li$|2\rangle$ spin mixture at a magnetic field near $1150\,$G on the attractive side of the $834$-G Li$|1\rangle$-Li$|2\rangle$ Feshbach resonance in a single-beam optical dipole trap \cite{Spiegelhalder2010aop}.
During evaporation, a few $10^{4}$ K atoms are sympathetically cooled by the Li environment.
The endpoint of evaporation is always set to the same final value.
After evaporation, we follow the scheme described in Ref.~\cite{Spiegelhalder2010aop} to transfer the atoms into a crossed-beam optical dipole trap and reach a magnetic field of $154.8\,$G with typically $10^6$ Li atoms in state Li$|1\rangle$ and $4 \times 10^4$ K atoms in state K$|1\rangle$.
We finally vary the temperature of our sample by increasing the power of our crossed beams to adiabatically recompress the trapped sample.
This scheme allows us to maintain a similar population imbalance and degeneracy for the three trap settings used.

To prepare for dimer association, we first create a weakly interacting Li$|1\rangle$-K$|3\rangle$ mixture at $B_0 +180\,$mG.
A first rf pulse transfers $\sim$80$\%$ of the K$|1\rangle$ population into state K$|2\rangle$ and a second rf pulse then transfers the total K$|2\rangle$ population into the interacting state K$|3\rangle$. 
The $\sim$7\,000 K atoms, which remain in the K$|1\rangle$ state, later serve for the spectroscopy.

We associate dimers using a two-step magnetic field ramp.
In a first 20-ms step we ramp the magnetic field from $B_0+180\,$mG to $B_0+5\,$mG.
This ramp is sufficiently slow for the Li atoms to be attracted into the regions of high K density, increasing the density overlap between the two clouds.
We then associate the Li$|1\rangle$K$|3\rangle$ dimers via a $0.5$-ms Feshbach ramp to $B_0-17\,$mG. 
We note that, during these magnetic field ramps, two-body inelastic losses \cite{Naik2011fri} are negligible. 

To obtain a pure sample of about 15\,000 Li$|1\rangle$K$|3\rangle$ dimers, we remove all unbound atoms from the states Li$|1\rangle$ and K$|3\rangle$.
The Li$|1\rangle$ atoms are removed by a sequence of rf and laser pulses. 
This procedure consists of a first 250-$\mu$s rf pulse resonant with the free Li$|1\rangle$$\rightarrow$Li$|2\rangle$ transition, followed by a 10-$\mu$s resonant light pulse, which selectively removes the Li$|2\rangle$ atoms from the trap. 
This scheme removes about 95$\%$ of the excess Li atoms without causing any observable loss of KLi dimers. 
A second 250-$\mu$s rf pulse transfers the leftover 5$\%$ of Li$|1\rangle$ atoms into the noninteracting Li$|2\rangle$ state, where they remain without further affecting the experiment. 

Simultaneously with this ``double-cleaning" of the unbound Li atoms, we remove the unbound K$|3\rangle$ atoms in a similar way. 
Using a 90-$\mu$s rf pulse resonant with the K$|3\rangle$$\rightarrow$K$|2\rangle$ transition, followed by a second 145-$\mu$s rf pulse resonant with the K$|3\rangle$$\rightarrow$K$|4\rangle$ transition, we empty the K$|3\rangle$ state with $>$99\,\% efficiency. 
The pulse lengths are chosen such that they are short, i.e. spectroscopically wide, compared to the frequency shifts due to atom-dimer and atom-atom interactions but long, i.e spectroscopically narrow, compared to the binding energy $E_{\rm b} = h \times 17\,$kHz ($h$ is Planck's constant) of the dimers, avoiding the dissociation of dimers.

In a final step, the $\sim$7\,000 K atoms which resided in state K$|1\rangle$ during the entire dimer association process, are transferred in the K$|2\rangle$ state and thus prepared for the rf spectroscopy.
This is accomplished by a rf pulse which flips the K$|1\rangle$ and K$|2\rangle$ populations.
We note that these K atoms remain unaffected by the dimer association since their interactions with the other components are negligible over the timescales of the experiment.

From here, we reach the specific magnetic field detunings $B-B_0$, at which the spectroscopy is performed, by a 200-$\mu$s magnetic field ramp.

\begin{center}
\begin{figure}
\includegraphics[width=0.9 \columnwidth, clip]{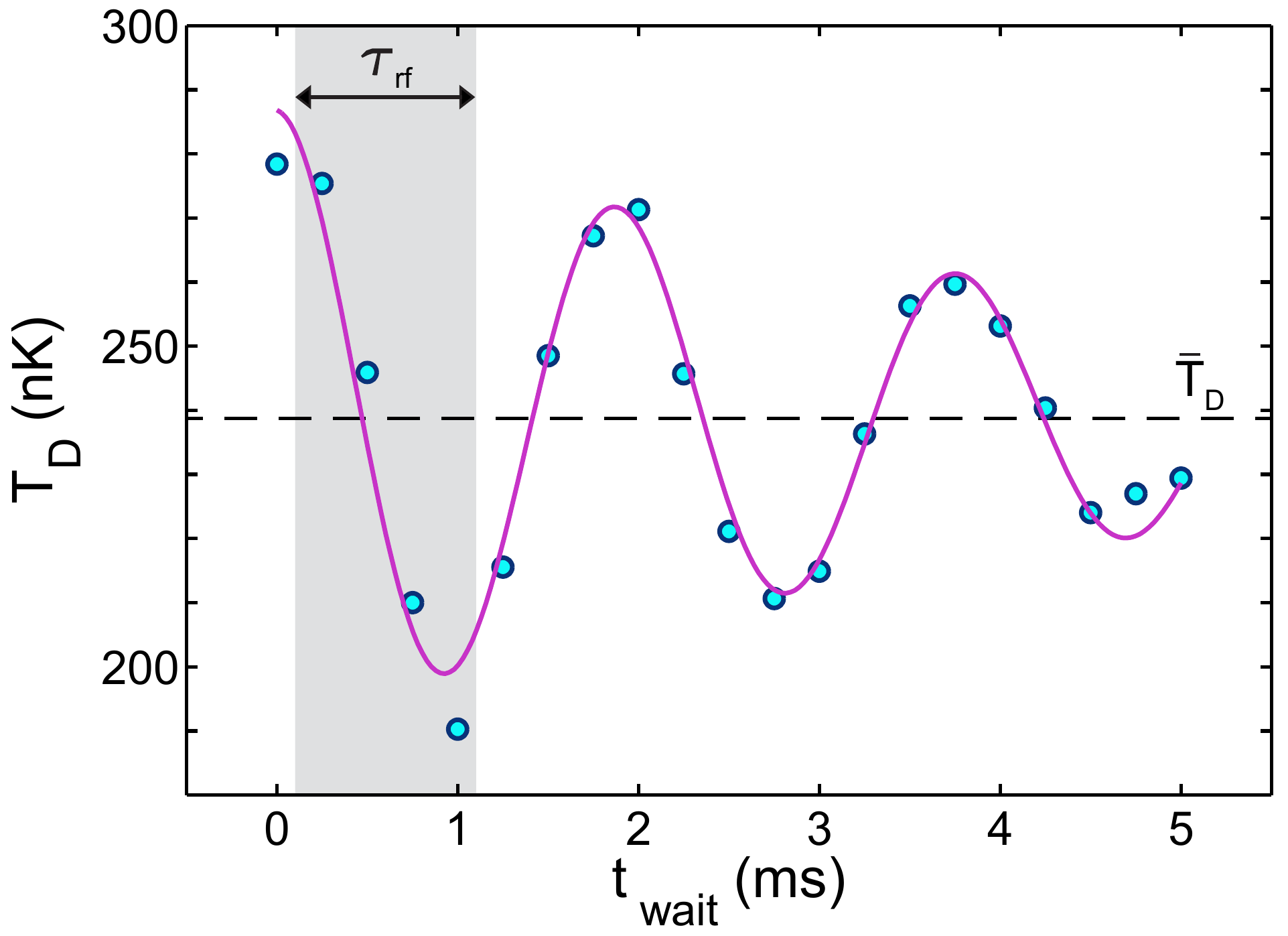}
\caption{Radial oscillation of the dimer cloud after the magnetic field ramp and the removal of the Li atoms. 
We plot the dimer temperature $T_{\rm D}$ versus the wait time $t_{\rm wait}$ after the first rf cleaning pulse to release from the trap. 
The filled circles are the experimental data, the solid line is a fit of a damped harmonic oscillation to the data. 
The shaded area indicates the time at which the spectroscopy rf pulses are applied and the dashed line marks the experimentally relevant averaged dimer temperature $\bar{T}_{D}$.}
\label{FigureSI3}
\end{figure}
\end{center}

\vspace{-1.3cm}
\subsection{Determination of the temperatures and the densities}

Here, we describe how we determine the temperatures and the densities of the atom cloud and the dimer cloud.
The resulting experimental parameters are summarized in Table~\ref{Table1}.

\begin{table*}
\begin{tabular}{|c|c|c|c|c|c|c|c|c|c|c|c|c|c|}
  \hline
  Trap & $T_{\rm eff}$	& $T_{\rm K}$	& $\bar{T}_{\rm D}$	& $\nu_{r, {\rm  K}}$	&	$\nu_{ a, {\rm K}}$ 	& $\nu_{ r, {\rm Li}}$	& $\nu_{a, {\rm Li}}$ & $\nu_{r, {\rm D}}$ & $\nu_{a, {\rm D}}$ & $\sigma_{r, {\rm K}}$	& $\sigma_{a, {\rm K}}$ & $\bar{\sigma}_{r, {\rm D}}$	& $\sigma_{a, {\rm D}}$	\\
   & (nK)		& (nK)				& (nK)				& (Hz)						&	(Hz)							& (Hz)							& (Hz) 						 & (Hz) 					 & (Hz) 				  	& ($\mu$m) 				  & ($\mu$m) 				   & 
($\mu$m) 				   & ($\mu$m) 				  \\
  \hline
  \hline
  1 &165(15) &138(5) &195(15) &197(5) &25.5(10) &314(5) &34.0(10) &216(5) &27.0(10) &4.3(1) &33(2) &4.4(1) &36(2)\\
    \hline
  2 &232(15) &225(5) &240(15) &284(5) &36.4(10) &446(5) &54.6(10) &310(5) &39.3(10) &3.8(1) &30(2) &3.4(1) &33(2)\\
    \hline
  3 &370(15) &345(5) &398(15) &415(5) &54.0(10) &671(5) &85.0(10) &457(5) &59.0(10) &3.2(1) &25(2) &2.9(1) &26(2)\\
  \hline
\end{tabular}
\caption{Parameters characterizing the three exploited trap settings. 
The table shows the effectice atom-dimer temperature $T_{\rm eff}$, the temperature of the K atoms, $T_{\rm K}$, and the average dimer temperature, $\bar{T}_{\rm D}$. 
From the radial (axial) trap frequencies of K and Li, $\nu_{r(a),\rm K}$ and $\nu_{r(a),\rm Li}$, we determine the trap frequencies $\nu_{r(a),\rm D}$ of the dimers. 
We also show the axial and radial in-situ Gaussian widths of dimers (K atoms), $\sigma_{a,\rm D(K)}$ and  $\sigma_{r,\rm D(K)}$, respectively.}
\label{Table1}
\end{table*}

{\em Atom and dimer temperatures} -- 
The temperatures of our atom and dimer clouds are obtained by Gaussian fits to absorption images of the clouds after a long time-of-flight of $t_{\rm tof} = 6\,{\rm ms}$.
With the measured radial Gaussian width $\sigma_{\rm tof, K(D)}$ the atom (dimer) temperature $T_{\rm K(D)}$ is given by
\begin{equation}
k_{\rm B}T_{\rm K(D)} =   m_{\rm K(D)} \left(\sigma_{\rm tof, K(D)}/t_{\rm tof}\right)^2,
\label{TempDeterm}
\end{equation}
where $m_{\rm K(D)}$ is the mass of the atom (dimer).

The magnetic field ramps and the removal of the surrounding Li shell, described in the previous section, excite collective oscillations of the dimer cloud. 
We trace these oscillations in momentum space as a function of a wait time $t_{\rm wait}$ after the cleaning procedure to release from the trap.
An example of such an oscillation is shown in Fig.~\ref{FigureSI3}.
In order to characterize the temperature at the time of the experiment, i.e.\ during the application of the 1-ms rf pulse (shaded area), we introduce the average temperature
\begin{equation}
\bar{T}_{\rm D} = \frac{1}{\tau_{\rm rf}}\int \limits_{\rm rf} {T_{\rm D} dt}.
\label{AvgTemp} 
\end{equation}

{\em Axial and radial sizes} --
To determine the densities of the atom (K) cloud and the dimer (D) cloud, we measure their Gaussian radial ($r$) and axial ($a$) widths $\sigma_{r,\rm K(D)}$ and $\sigma_{a,\rm K(D)}$, respectively.
The axial widths are measured from a Gaussian fit to the axial profiles of in-situ absorption images.
Since the radial widths are on the order of our imaging resolution, they can not be determined from in-situ images.
We instead determine the radial widths of the K atom cloud as
\begin{equation}
\sigma_{r, \rm K} = \sqrt{\frac{k_{\rm B}T_{\rm K}}{m_{\rm K} (2\pi\nu_{r,\rm K})^{2}}},
\label{SigmaRadial}
\end{equation}
where $T_{\rm K}$ and $\nu_{r,\rm K}$ denote the temperature and the radial trap frequency of the K atoms, respectively.
Accordingly we determine the average radial in-situ width of the dimers,
\begin{equation}
\bar{\sigma}_{r, \rm D} = \sqrt{\frac{k_{\rm B}\bar{T}_{\rm D}}{m_{\rm D} (2\pi\nu_{r,\rm D})^{2}}},
\label{SigmaRadDimers}
\end{equation}
using the averaged dimer temperature $\bar{T}_{\rm D}$, and the radial dimer trap frequency $\nu_{r,\rm D}$.

{\em Trap frequencies of the dimers} -- 
We use the measured trap frequencies of the K and Li atoms to determine the trap frequencies $\nu_{r(a),\rm D}$of the LiK-dimers.
Since the dimers are weakly bound over the magnetic field range investigated, their polarizabilities are approximately given by the sum of the polarizabilities of the Li and the K atoms.
We want to point out that the differential light shift, shifting the FR center (see section \ref{lightshift}), gives only a $<10\%$ correction to the trap potential and is neglected.
Therefore, to a good aproximation, the dimer trap frequencies are given by
\begin{equation}
\nu_{a(r), \rm D}= \sqrt{(m_{\rm K} \nu_{a(r),\rm K}^{2} + m_{\rm Li} \nu_{a(r),\rm Li}^{2})/m_{\rm D}},
\label{moltrapfrequ}
\end{equation}
with $m_{\rm Li}$ being the mass of a Li atom.

{\em Mean dimer density} -- 
For a given dimer number, $N_{\rm D}$, the mean dimer density experienced by the K atoms $\bar{n}_{\rm D}$ is given by 
\begin{equation}
\bar{n}_{\rm D} = \frac{N_{\rm D}} {(2\pi)^{3/2} (\sigma_{r, \rm K}^2 + \bar{\sigma}_{r, \rm D}^2) \sqrt{\sigma_{a, \rm K}^2 + \sigma_{a, \rm D}^2}},
\end{equation}
where we have assumed Gaussian-shaped atom and dimer clouds.

{\em Effective temperature} -- 
Due to heating and oscillations caused by our preparation procedure, the dimer temperature $T_{\rm D}$ in our system is different from the temperature of the non-interacting K$|2\rangle$ atoms that we use for rf spectroscopy. 
However, since our dimer and atom clouds are both non-degenerate, the energies of the atom-dimer collisions still assume a Boltzmann distribution. 
Averaging this distribution over the oscillations of the dimer cloud results in an effective atom-dimer collision temperature
\begin{equation}
 T_{\rm eff} = \mu_3 (T_{\rm K}/m_{\rm K} + \bar{T}_{\rm D}/m_{\rm D}),
\label{efftempMIX}
\end{equation}
where $\mu_3 = m_{\rm K}m_{\rm D}/(m_{\rm K}+m_{\rm D})$ is the atom-dimer reduced mass.

\vspace{-0.2cm}
\subsection{Importance of higher partial wave scattering and comparison to the equal-mass case}

\begin{center}
\begin{figure}[H]
\includegraphics[width=0.95 \columnwidth, clip]{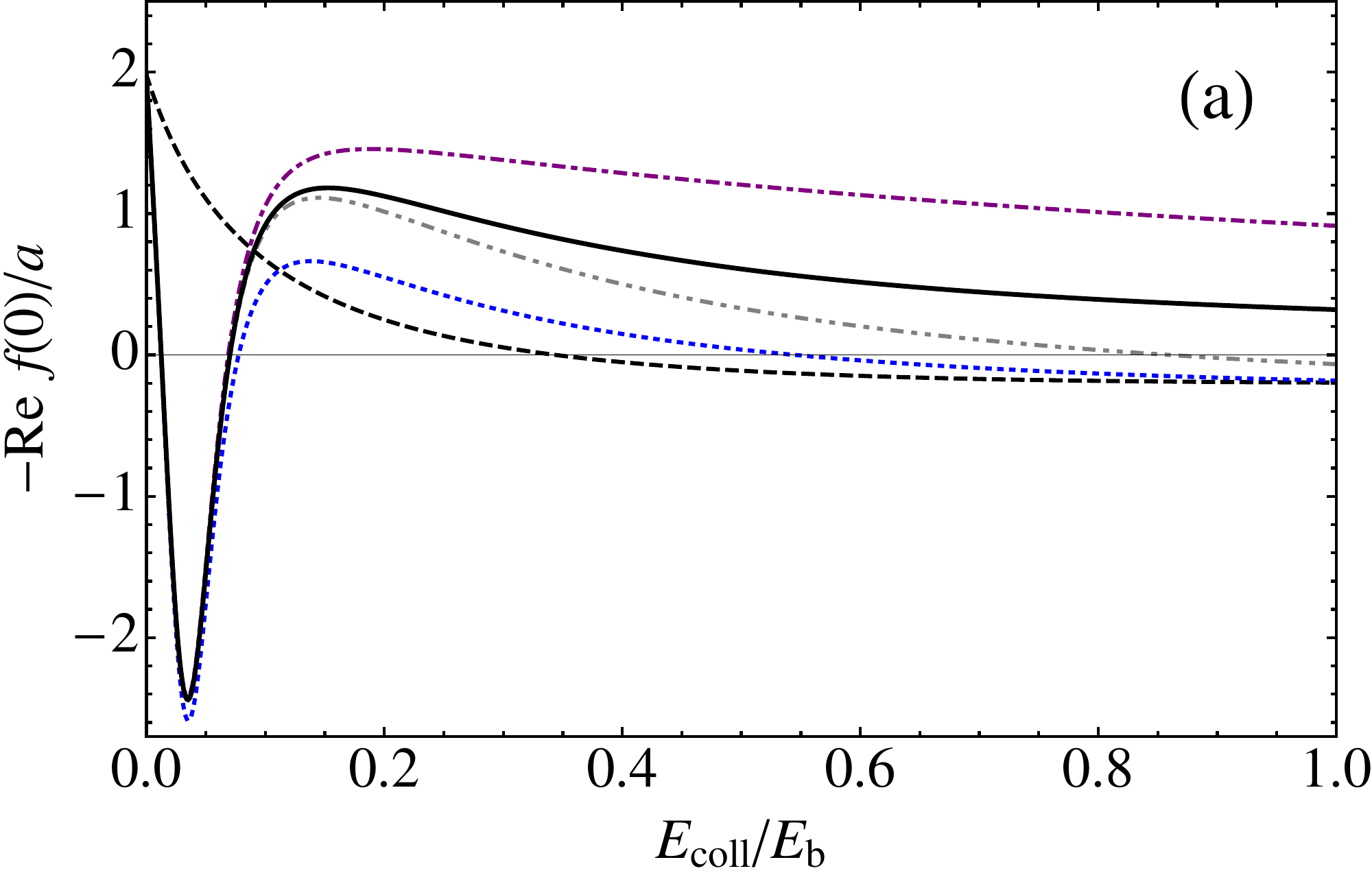}
\vskip 10pt
\includegraphics[width=0.95 \columnwidth, clip]{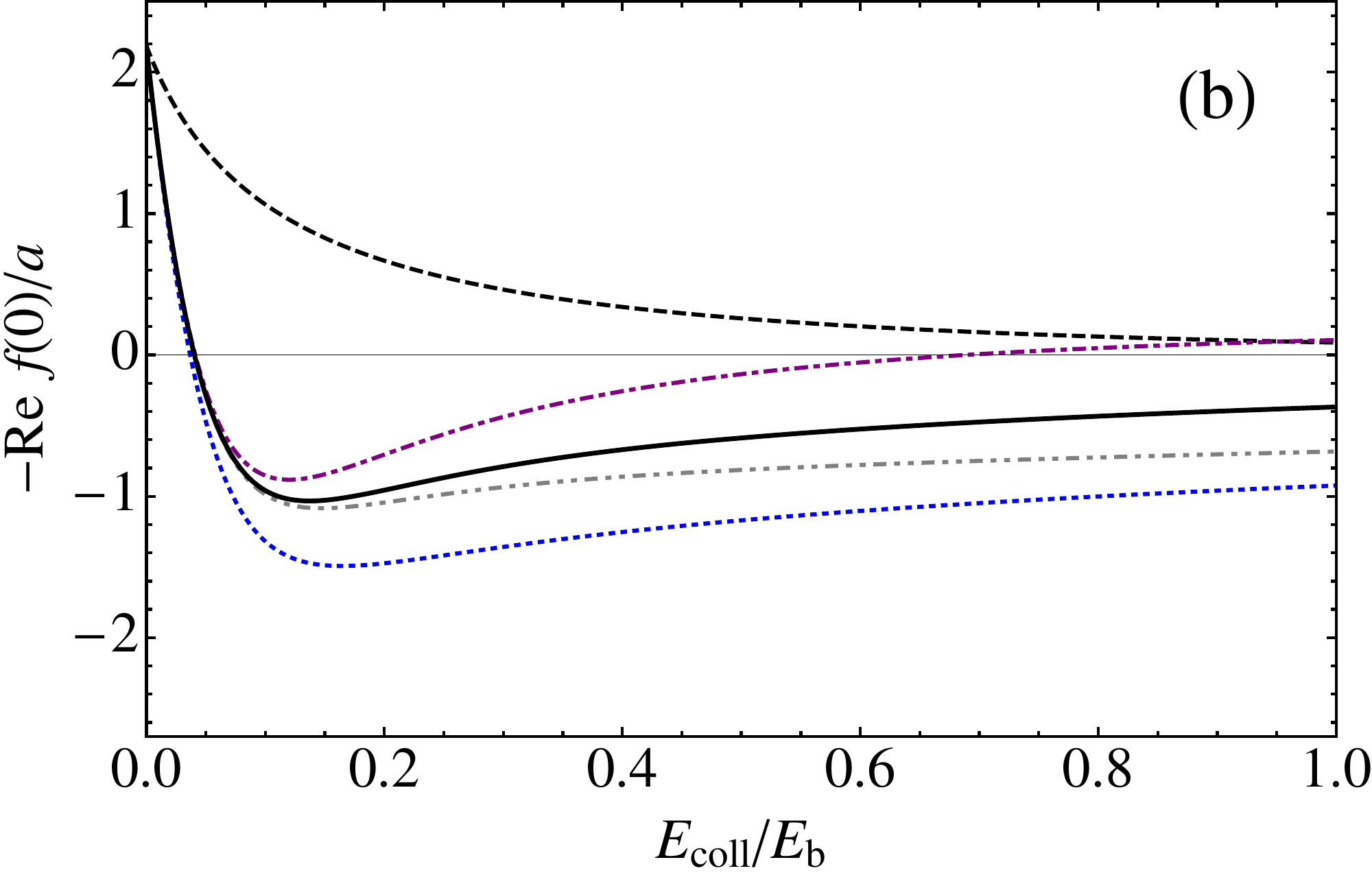}
\vskip 10pt
\includegraphics[width=0.95 \columnwidth, clip]{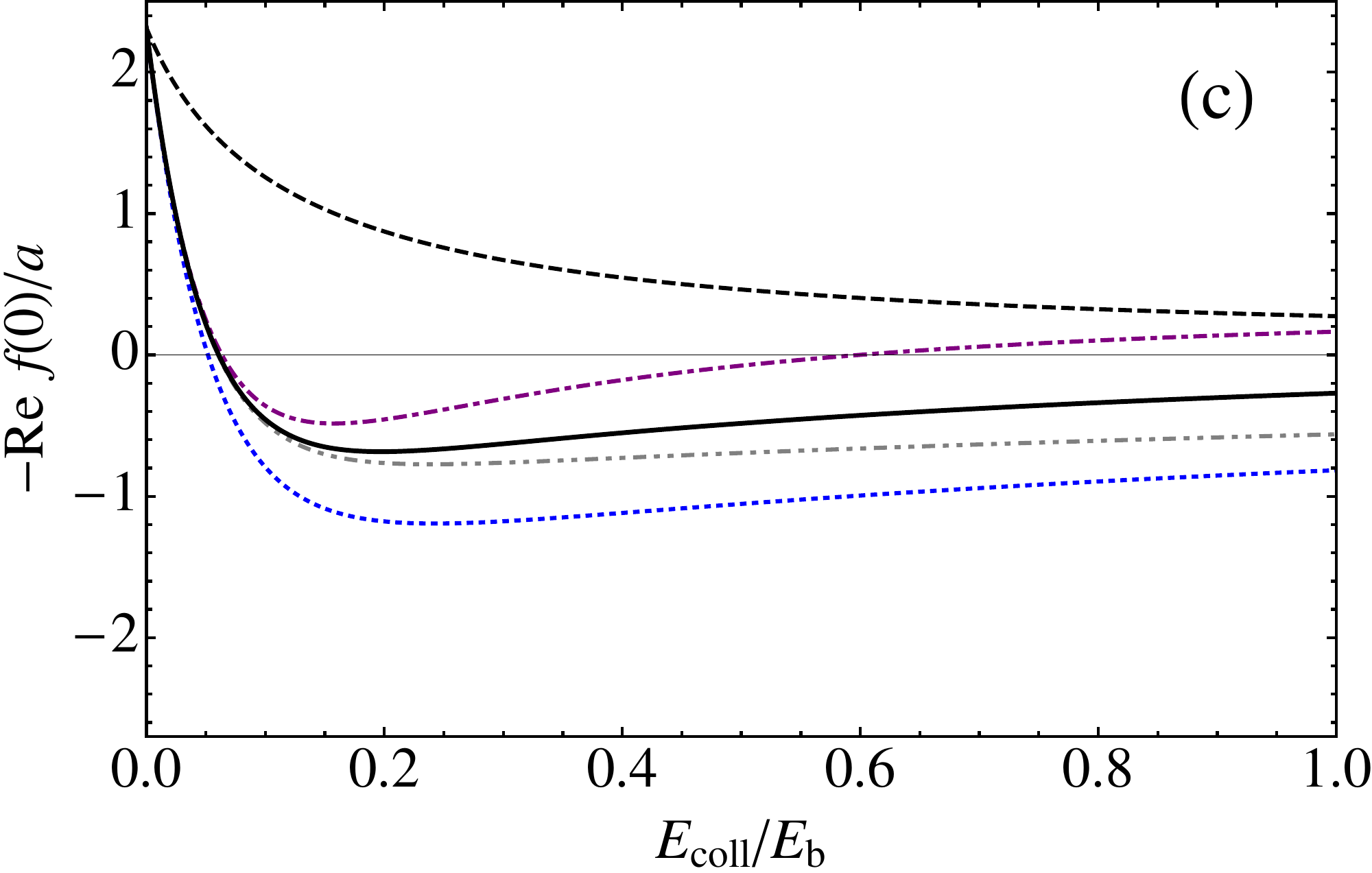}
\caption{Scattering of a $^{40}$K atom with a $^{6}$Li$^{40}$K dimer.
The quantitiy $-\mbox{Re}\,f(0)$ is plotted as a function of atom-dimer collision energy for (a) $R^*/a=0$ [$B-B_0=0$], (b) $R^*/a=1/2$  [$B-B_0=-10\,$mG], and (c) $R^*/a=1$  [$B-B_0=-21\,$mG]. 
The lines are including $s$-wave scattering only (black, dashed), including up to $p$-wave (blue, dotted), up to $d$-wave (purple, dot-dashed), and up to $f$-wave (gray, double dot-dashed). 
The solid black line is $-\mbox{Re}\,f(0)$ including the first 16 partial waves.}
\label{FigureSI4}
\end{figure}
\end{center}

In this Section, we justify several important statements made in the main text.
First, we have argued that the range of the atom-dimer interaction is comparable with the typical de Broglie wavelength and, therefore, quite a few partial waves are necessary to quantitatively characterize the line shift. 
In Fig.~\ref{FigureSI4}, we display $-\mbox{Re}\,f(0)$, the quantity which is thermally averaged in the main text to obtain the line shifts. 
The method of calculating the scattering amplitude is described in Ref.~\cite{Levinsen2009ads}.
Remarkably, the real part of the forward-scattering amplitude is seen to change sign at a collision energy much smaller than the binding energy, even for a relatively large detuning of 21$\,$mG.
The second change of sign of $-\mbox{Re}\,f(0)$ seen in Fig.~\ref{FigureSI4}(a) is attributed to the fact that $\delta_p$ exceeds $\pi/2$ above $E_{\mbox{\small coll}} \approx 0.1 E_{\mbox{\small b}}$, the point of the $p$-wave resonance. 
The $p$-wave contribution at larger collision energies then becomes positive (repulsive) [see Eq.~(1) of the main text]. 
However, this peculiar phenomenon takes place only in a very close vicinity of the wide resonance limit as the $p$-wave phase shift drops rather abruptly with $R^*/a$ \cite{Levinsen2009ads}.
We also note how, as the collision energy is increased, more and more partial wave channels are needed to accurately describe the forward-scattering amplitude. 
The calculation presented here includes the first 16 partial waves, which is sufficient to obtain an essentially converged scattering amplitude at the dimer breakup threshold.

\begin{center}
\begin{figure}[H]
\includegraphics[width=0.95 \columnwidth, clip]{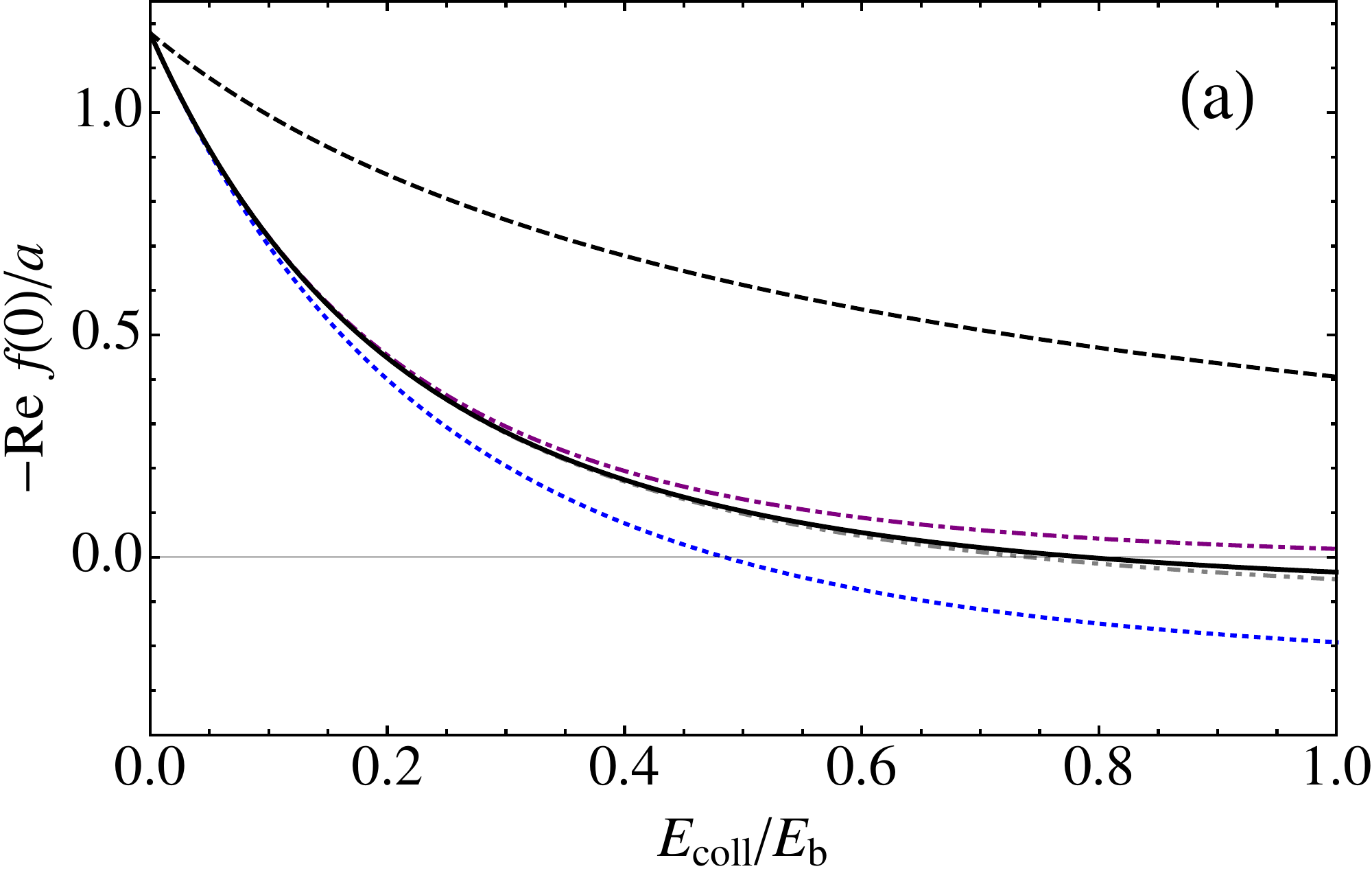}
\vskip 10pt
\includegraphics[width=0.95 \columnwidth, clip]{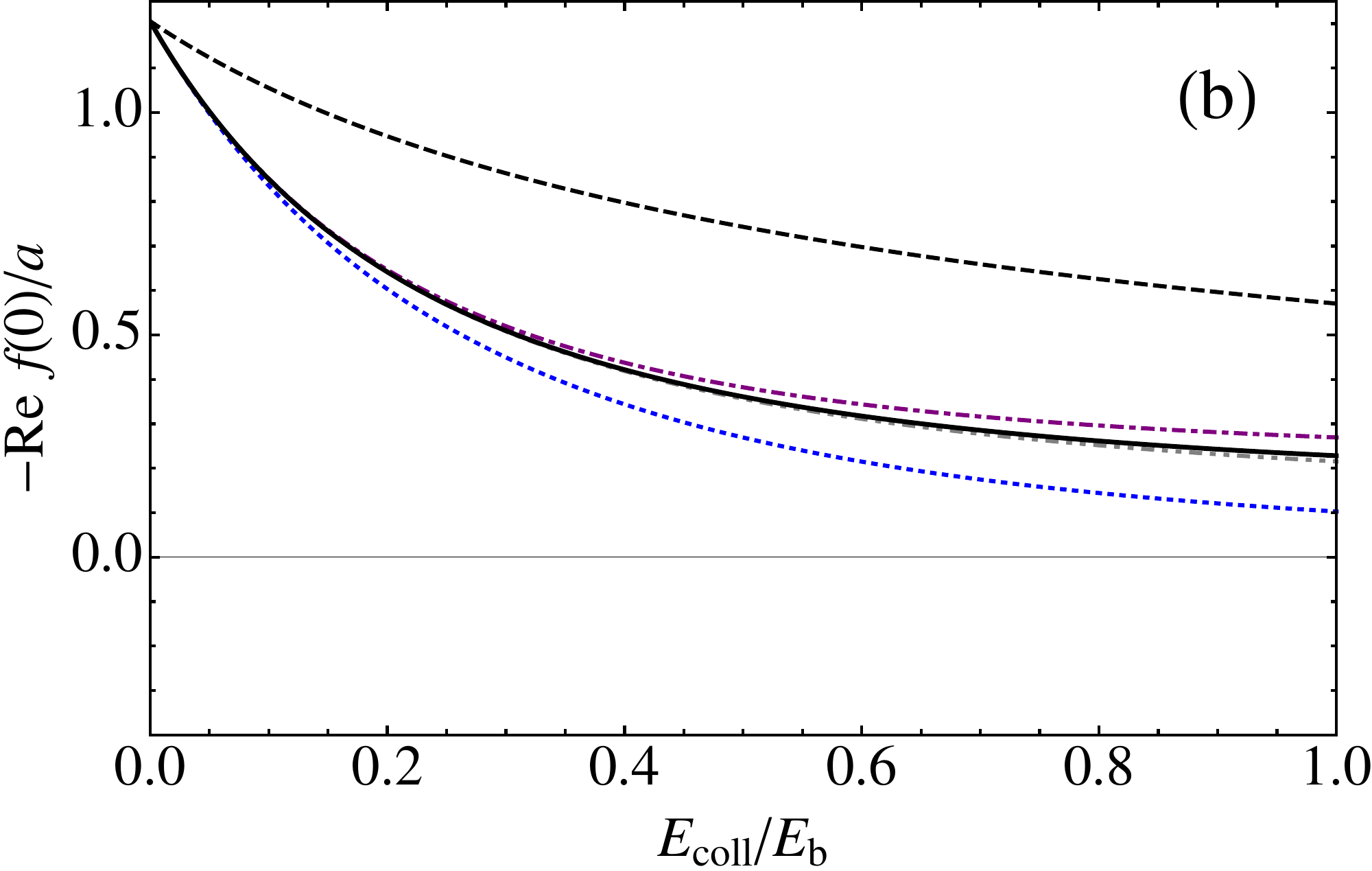}
\vskip 10pt
\includegraphics[width=0.95 \columnwidth, clip]{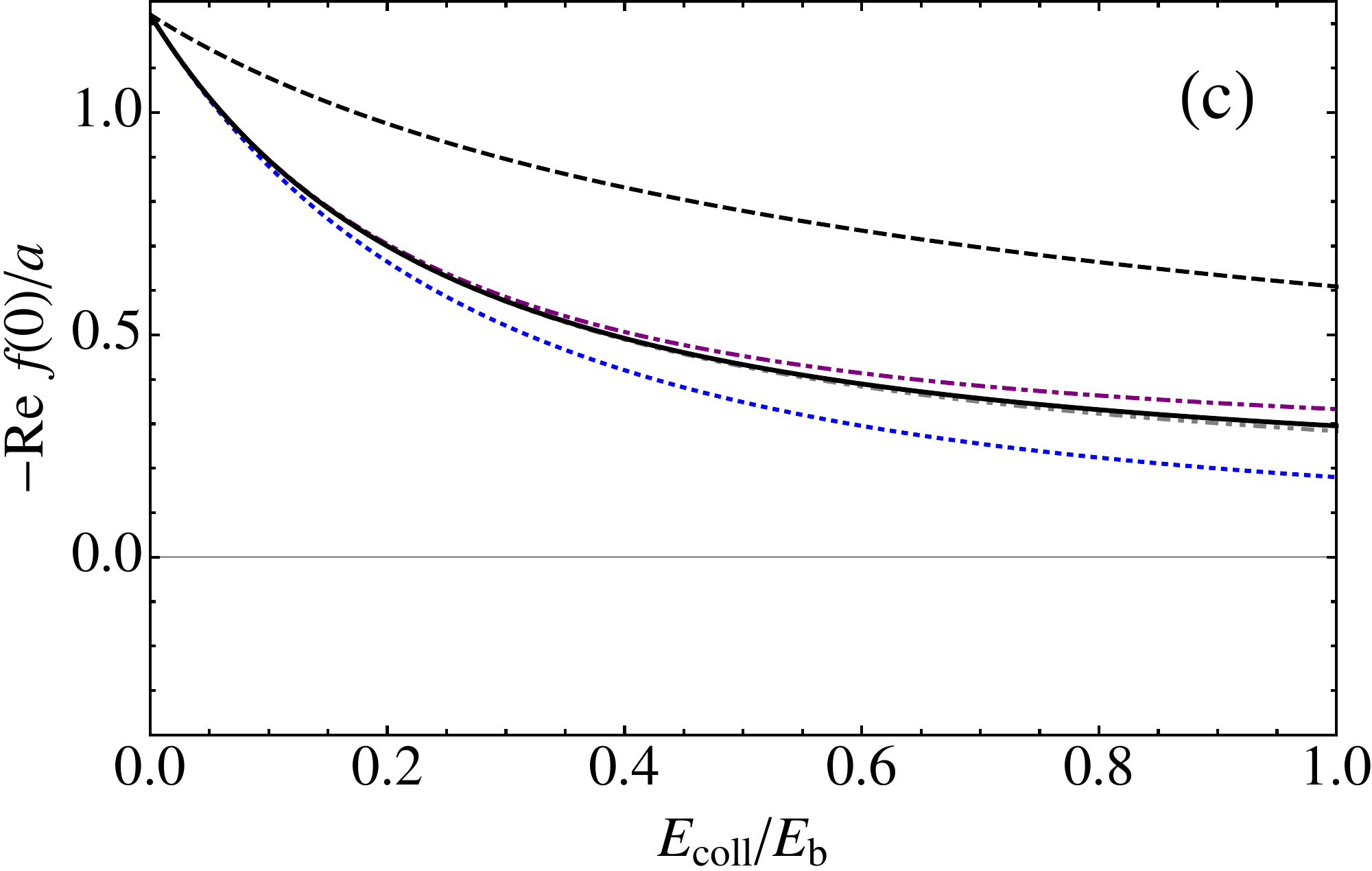}
\caption{Equal-mass case of atom-dimer scattering.
We plot $-\mbox{Re}\,f(0)$ as a function of collision energy for the homonuclear case, $m_\uparrow=m_\downarrow$. 
The conventions used for the lines as well as the detunings in (a) to (c) are the same as in Fig.~\ref{FigureSI4}. 
The solid black line includes the first 9 partial waves.}
\label{FigureSI5}
\end{figure}
\end{center}

As far as the equal mass case is concerned, the competition between the attraction in odd partial waves and repulsion in even partial waves is also quite significant, yet much less pronounced compared to the K-Li case. 
In Fig.~\ref{FigureSI5} we display $-\mbox{Re}\,f(0)$ as a function of $E_{\mbox{\small coll}}$ for equal masses. 
Here the broad resonance case in Fig.~\ref{FigureSI5}(a) is relevant since it is readily available in current experiments and since there the effect of higher partial waves is most noticeable. 
We see that the forward-scattering amplitude does change sign in this case. 
However, in contrast to the K-Li case, this happens at a high collision energy close to the dimer breakup threshold and, in fact, already for $R^*/a\gtrsim0.03$ the crossing is no longer on the scale. Thus, in the narrow resonance case illustrated in Fig.~\ref{FigureSI5}(b) and (c) the interaction is found to be repulsive below the dimer breakup threshold.
In all cases the thermally averaged quantity $-\mbox{Re}\,\langle f(0) \rangle$ is positive.

Finally, let us also make a remark concerning the thermal averaging of the scattering amplitude which we use in the main text. 
In principle, the averaging procedure requires the knowledge of the phase shifts above the atom-dimer breakup threshold. 
However, we always restrict ourselves to temperatures $k_{\rm B}T\lesssim E_{\mbox{\small b}}/2$ and we check that in this case the integration result is insensitive to the exact extrapolation scheme. 
In practice we extrapolate the phase shift $\delta_l(k)$ using the log function, which works very well when we calculate the phase shifts above the breakup threshold in the Born-Oppenheimer approximation~\cite{Levinsen2011ada}.

\bibliographystyle{apsrev}

\begin{thebibliography}{40}
\expandafter\ifx\csname natexlab\endcsname\relax\def\natexlab#1{#1}\fi
\expandafter\ifx\csname bibnamefont\endcsname\relax
  \def\bibnamefont#1{#1}\fi
\expandafter\ifx\csname bibfnamefont\endcsname\relax
  \def\bibfnamefont#1{#1}\fi
\expandafter\ifx\csname citenamefont\endcsname\relax
  \def\citenamefont#1{#1}\fi
\expandafter\ifx\csname url\endcsname\relax
  \def\url#1{\texttt{#1}}\fi
\expandafter\ifx\csname urlprefix\endcsname\relax\def\urlprefix{URL }\fi
\providecommand{\bibinfo}[2]{#2}
\providecommand{\eprint}[2][]{\url{#2}}

\bibitem[{\citenamefont{Giorgini et~al.}(2008)\citenamefont{Giorgini,
  Pitaevskii, and Stringari}}]{Giorgini2008tou}
\bibinfo{author}{\bibfnamefont{S.}~\bibnamefont{Giorgini}},
  \bibinfo{author}{\bibfnamefont{L.~P.} \bibnamefont{Pitaevskii}},
  \bibnamefont{and}
  \bibinfo{author}{\bibfnamefont{S.}~\bibnamefont{Stringari}},
  \bibinfo{journal}{Rev. Mod. Phys.} \textbf{\bibinfo{volume}{80}},
  \bibinfo{pages}{1215} (\bibinfo{year}{2008}).

\bibitem[{\citenamefont{Bloch et~al.}(2008)\citenamefont{Bloch, Dalibard, and
  Zwerger}}]{Bloch2008mbp}
\bibinfo{author}{\bibfnamefont{I.}~\bibnamefont{Bloch}},
  \bibinfo{author}{\bibfnamefont{J.}~\bibnamefont{Dalibard}}, \bibnamefont{and}
  \bibinfo{author}{\bibfnamefont{W.}~\bibnamefont{Zwerger}},
  \bibinfo{journal}{Rev. Mod. Phys.} \textbf{\bibinfo{volume}{80}},
  \bibinfo{pages}{885} (\bibinfo{year}{2008}).

\bibitem[{\citenamefont{Chin et~al.}(2010)\citenamefont{Chin, Grimm, Julienne,
  and Tiesinga}}]{Chin2010fri}
\bibinfo{author}{\bibfnamefont{C.}~\bibnamefont{Chin}},
  \bibinfo{author}{\bibfnamefont{R.}~\bibnamefont{Grimm}},
  \bibinfo{author}{\bibfnamefont{P.~S.} \bibnamefont{Julienne}},
  \bibnamefont{and} \bibinfo{author}{\bibfnamefont{E.}~\bibnamefont{Tiesinga}},
  \bibinfo{journal}{Rev. Mod. Phys.} \textbf{\bibinfo{volume}{82}},
  \bibinfo{pages}{1225} (\bibinfo{year}{2010}).

\bibitem[{\citenamefont{Iskin and {S\'{a} de Melo}}(2006)}]{Iskin2006tsf}
\bibinfo{author}{\bibfnamefont{M.}~\bibnamefont{Iskin}} \bibnamefont{and}
  \bibinfo{author}{\bibfnamefont{C.~A.~R.} \bibnamefont{{S\'{a} de Melo}}},
  \bibinfo{journal}{Phys. Rev. Lett.} \textbf{\bibinfo{volume}{97}},
  \bibinfo{eid}{100404} (\bibinfo{year}{2006}).

\bibitem[{\citenamefont{Bausmerth et~al.}(2009)\citenamefont{Bausmerth, Recati,
  and Stringari}}]{Bausmerth2009ccl}
\bibinfo{author}{\bibfnamefont{I.}~\bibnamefont{Bausmerth}},
  \bibinfo{author}{\bibfnamefont{A.}~\bibnamefont{Recati}}, \bibnamefont{and}
  \bibinfo{author}{\bibfnamefont{S.}~\bibnamefont{Stringari}},
  \bibinfo{journal}{Phys. Rev. A} \textbf{\bibinfo{volume}{79}},
  \bibinfo{pages}{043622} (\bibinfo{year}{2009}).

\bibitem[{\citenamefont{Gezerlis et~al.}(2009)\citenamefont{Gezerlis, Gandolfi,
  Schmidt, and Carlson}}]{Gezerlis2009hlf}
\bibinfo{author}{\bibfnamefont{A.}~\bibnamefont{Gezerlis}},
  \bibinfo{author}{\bibfnamefont{S.}~\bibnamefont{Gandolfi}},
  \bibinfo{author}{\bibfnamefont{K.~E.} \bibnamefont{Schmidt}},
  \bibnamefont{and} \bibinfo{author}{\bibfnamefont{J.}~\bibnamefont{Carlson}},
  \bibinfo{journal}{Phys. Rev. Lett.} \textbf{\bibinfo{volume}{103}},
  \bibinfo{pages}{060403} (\bibinfo{year}{2009}).

\bibitem[{\citenamefont{von Keyserlingk and
  Conduit}(2011)}]{Keyserlingk2011ifi}
\bibinfo{author}{\bibfnamefont{C.~W.} \bibnamefont{von Keyserlingk}}
  \bibnamefont{and} \bibinfo{author}{\bibfnamefont{G.~J.}
  \bibnamefont{Conduit}}, \bibinfo{journal}{Phys. Rev. A}
  \textbf{\bibinfo{volume}{83}}, \bibinfo{pages}{053625}
  (\bibinfo{year}{2011}).

\bibitem[{\citenamefont{Sotnikov et~al.}(2012)\citenamefont{Sotnikov, Cocks,
  and Hofstetter}}]{Sotnikov2012aom}
\bibinfo{author}{\bibfnamefont{A.}~\bibnamefont{Sotnikov}},
  \bibinfo{author}{\bibfnamefont{D.}~\bibnamefont{Cocks}}, \bibnamefont{and}
  \bibinfo{author}{\bibfnamefont{W.}~\bibnamefont{Hofstetter}},
  \bibinfo{journal}{Phys. Rev. Lett.} \textbf{\bibinfo{volume}{109}},
  \bibinfo{pages}{065301} (\bibinfo{year}{2012}).

\bibitem[{\citenamefont{Cui and Ho}(2013)}]{Cui2013psi}
\bibinfo{author}{\bibfnamefont{X.}~\bibnamefont{Cui}} \bibnamefont{and}
  \bibinfo{author}{\bibfnamefont{T.-L.} \bibnamefont{Ho}},
  \bibinfo{journal}{Phys. Rev. Lett.} \textbf{\bibinfo{volume}{110}},
  \bibinfo{pages}{165302} (\bibinfo{year}{2013}).

\bibitem[{\citenamefont{Gubbels et~al.}(2009)\citenamefont{Gubbels, Baarsma,
  and Stoof}}]{Gubbels2009lpi}
\bibinfo{author}{\bibfnamefont{K.~B.} \bibnamefont{Gubbels}},
  \bibinfo{author}{\bibfnamefont{J.~E.} \bibnamefont{Baarsma}},
  \bibnamefont{and} \bibinfo{author}{\bibfnamefont{H.~T.~C.}
  \bibnamefont{Stoof}}, \bibinfo{journal}{Phys. Rev. Lett.}
  \textbf{\bibinfo{volume}{103}}, \bibinfo{pages}{195301}
  (\bibinfo{year}{2009}).

\bibitem[{\citenamefont{Mathy et~al.}(2011)\citenamefont{Mathy, Parish, and
  Huse}}]{Mathy2011tma}
\bibinfo{author}{\bibfnamefont{C.~J.~M.} \bibnamefont{Mathy}},
  \bibinfo{author}{\bibfnamefont{M.~M.} \bibnamefont{Parish}},
  \bibnamefont{and} \bibinfo{author}{\bibfnamefont{D.~A.} \bibnamefont{Huse}},
  \bibinfo{journal}{Phys. Rev. Lett.} \textbf{\bibinfo{volume}{106}},
  \bibinfo{pages}{166404} (\bibinfo{year}{2011}).

\bibitem[{\citenamefont{Qi and Zhai}(2012)}]{Qi2012hpf}
\bibinfo{author}{\bibfnamefont{R.}~\bibnamefont{Qi}} \bibnamefont{and}
  \bibinfo{author}{\bibfnamefont{H.}~\bibnamefont{Zhai}},
  \bibinfo{journal}{Phys. Rev. A} \textbf{\bibinfo{volume}{85}},
  \bibinfo{pages}{041603(R)} (\bibinfo{year}{2012}).

\bibitem[{\citenamefont{Daily and Blume}(2012)}]{Daily2012tot}
\bibinfo{author}{\bibfnamefont{K.~M.} \bibnamefont{Daily}} \bibnamefont{and}
  \bibinfo{author}{\bibfnamefont{D.}~\bibnamefont{Blume}},
  \bibinfo{journal}{Phys. Rev. A} \textbf{\bibinfo{volume}{85}},
  \bibinfo{pages}{013609} (\bibinfo{year}{2012}).

\bibitem[{\citenamefont{Petrov et~al.}(2007)\citenamefont{Petrov,
  Astrakharchik, Papoular, Salomon, and Shlyapnikov}}]{Petrov2007cpo}
\bibinfo{author}{\bibfnamefont{D.~S.} \bibnamefont{Petrov}},
  \bibinfo{author}{\bibfnamefont{G.~E.} \bibnamefont{Astrakharchik}},
  \bibinfo{author}{\bibfnamefont{D.~J.} \bibnamefont{Papoular}},
  \bibinfo{author}{\bibfnamefont{C.}~\bibnamefont{Salomon}}, \bibnamefont{and}
  \bibinfo{author}{\bibfnamefont{G.~V.} \bibnamefont{Shlyapnikov}},
  \bibinfo{journal}{Phys. Rev. Lett.} \textbf{\bibinfo{volume}{99}},
  \bibinfo{eid}{130407} (\bibinfo{year}{2007}).

\bibitem[{\citenamefont{Baranov et~al.}(2008)\citenamefont{Baranov, Lobo, and
  Shlyapnikov}}]{Baranov2008spb}
\bibinfo{author}{\bibfnamefont{M.~A.} \bibnamefont{Baranov}},
  \bibinfo{author}{\bibfnamefont{C.}~\bibnamefont{Lobo}}, \bibnamefont{and}
  \bibinfo{author}{\bibfnamefont{G.~V.} \bibnamefont{Shlyapnikov}},
  \bibinfo{journal}{Phys. Rev. A} \textbf{\bibinfo{volume}{78}},
  \bibinfo{pages}{033620} (\bibinfo{year}{2008}).

\bibitem[{\citenamefont{Sanchez-Castro and Bedell}(1991)}]{Sanchez1991tcf}
\bibinfo{author}{\bibfnamefont{C.}~\bibnamefont{Sanchez-Castro}}
  \bibnamefont{and} \bibinfo{author}{\bibfnamefont{K.~S.}
  \bibnamefont{Bedell}}, \bibinfo{journal}{Phys. Rev. B}
  \textbf{\bibinfo{volume}{43}}, \bibinfo{pages}{12874} (\bibinfo{year}{1991}).

\bibitem[{\citenamefont{Orso et~al.}(2010)\citenamefont{Orso, Burovski, and
  Jolicoeur}}]{Orso2010llo}
\bibinfo{author}{\bibfnamefont{G.}~\bibnamefont{Orso}},
  \bibinfo{author}{\bibfnamefont{E.}~\bibnamefont{Burovski}}, \bibnamefont{and}
  \bibinfo{author}{\bibfnamefont{T.}~\bibnamefont{Jolicoeur}},
  \bibinfo{journal}{Phys. Rev. Lett.} \textbf{\bibinfo{volume}{104}},
  \bibinfo{pages}{065301} (\bibinfo{year}{2010}).

\bibitem[{\citenamefont{Dalmonte et~al.}(2012)\citenamefont{Dalmonte,
  Dieckmann, Roscilde, Hartl, Feiguin, Schollw\"ock, and
  Heidrich-Meisner}}]{Dalmonte2012dta}
\bibinfo{author}{\bibfnamefont{M.}~\bibnamefont{Dalmonte}},
  \bibinfo{author}{\bibfnamefont{K.}~\bibnamefont{Dieckmann}},
  \bibinfo{author}{\bibfnamefont{T.}~\bibnamefont{Roscilde}},
  \bibinfo{author}{\bibfnamefont{C.}~\bibnamefont{Hartl}},
  \bibinfo{author}{\bibfnamefont{A.~E.} \bibnamefont{Feiguin}},
  \bibinfo{author}{\bibfnamefont{U.}~\bibnamefont{Schollw\"ock}},
  \bibnamefont{and}
  \bibinfo{author}{\bibfnamefont{F.}~\bibnamefont{Heidrich-Meisner}},
  \bibinfo{journal}{Phys. Rev. A} \textbf{\bibinfo{volume}{85}},
  \bibinfo{pages}{063608} (\bibinfo{year}{2012}).

\bibitem[{\citenamefont{Nishida and Tan}(2008)}]{Nishida2008ufg}
\bibinfo{author}{\bibfnamefont{Y.}~\bibnamefont{Nishida}} \bibnamefont{and}
  \bibinfo{author}{\bibfnamefont{S.}~\bibnamefont{Tan}},
  \bibinfo{journal}{Phys. Rev. Lett.} \textbf{\bibinfo{volume}{101}},
  \bibinfo{pages}{170401} (\bibinfo{year}{2008}).

\bibitem[{\citenamefont{Nishida and Tan}(2009)}]{Nishida2009cie}
\bibinfo{author}{\bibfnamefont{Y.}~\bibnamefont{Nishida}} \bibnamefont{and}
  \bibinfo{author}{\bibfnamefont{S.}~\bibnamefont{Tan}},
  \bibinfo{journal}{Phys. Rev. A} \textbf{\bibinfo{volume}{79}},
  \bibinfo{pages}{060701(R)} (\bibinfo{year}{2009}).

\bibitem[{\citenamefont{Kartavtsev and Malykh}(2007)}]{Kartavtsev2007let}
\bibinfo{author}{\bibfnamefont{O.~I.} \bibnamefont{Kartavtsev}}
  \bibnamefont{and} \bibinfo{author}{\bibfnamefont{A.~V.}
  \bibnamefont{Malykh}}, \bibinfo{journal}{J. Phys. B}
  \textbf{\bibinfo{volume}{40}}, \bibinfo{pages}{1429} (\bibinfo{year}{2007}).

\bibitem[{\citenamefont{Wille et~al.}(2008)\citenamefont{Wille, Spiegelhalder,
  Kerner, Naik, Trenkwalder, Hendl, Schreck, Grimm, Tiecke, Walraven
  et~al.}}]{Wille2008eau}
\bibinfo{author}{\bibfnamefont{E.}~\bibnamefont{Wille}},
  \bibinfo{author}{\bibfnamefont{F.~M.} \bibnamefont{Spiegelhalder}},
  \bibinfo{author}{\bibfnamefont{G.}~\bibnamefont{Kerner}},
  \bibinfo{author}{\bibfnamefont{D.}~\bibnamefont{Naik}},
  \bibinfo{author}{\bibfnamefont{A.}~\bibnamefont{Trenkwalder}},
  \bibinfo{author}{\bibfnamefont{G.}~\bibnamefont{Hendl}},
  \bibinfo{author}{\bibfnamefont{F.}~\bibnamefont{Schreck}},
  \bibinfo{author}{\bibfnamefont{R.}~\bibnamefont{Grimm}},
  \bibinfo{author}{\bibfnamefont{T.~G.} \bibnamefont{Tiecke}},
  \bibinfo{author}{\bibfnamefont{J.~T.~M.} \bibnamefont{Walraven}},
  \bibnamefont{et~al.}, \bibinfo{journal}{Phys. Rev. Lett.}
  \textbf{\bibinfo{volume}{100}}, \bibinfo{eid}{053201} (\bibinfo{year}{2008}).

\bibitem[{\citenamefont{Costa et~al.}(2010)\citenamefont{Costa, Brachmann,
  Voigt, Hahn, Taglieber, H\"{a}nsch, and Dieckmann}}]{Costa2010swi}
\bibinfo{author}{\bibfnamefont{L.}~\bibnamefont{Costa}},
  \bibinfo{author}{\bibfnamefont{J.}~\bibnamefont{Brachmann}},
  \bibinfo{author}{\bibfnamefont{A.-C.} \bibnamefont{Voigt}},
  \bibinfo{author}{\bibfnamefont{C.}~\bibnamefont{Hahn}},
  \bibinfo{author}{\bibfnamefont{M.}~\bibnamefont{Taglieber}},
  \bibinfo{author}{\bibfnamefont{T.~W.} \bibnamefont{H\"{a}nsch}},
  \bibnamefont{and}
  \bibinfo{author}{\bibfnamefont{K.}~\bibnamefont{Dieckmann}},
  \bibinfo{journal}{Phys. Rev. Lett.} \textbf{\bibinfo{volume}{105}},
  \bibinfo{pages}{123201} (\bibinfo{year}{2010}).

\bibitem[{\citenamefont{Trenkwalder et~al.}(2011)\citenamefont{Trenkwalder,
  Kohstall, Zaccanti, Naik, Sidorov, Schreck, and Grimm}}]{Trenkwalder2011heo}
\bibinfo{author}{\bibfnamefont{A.}~\bibnamefont{Trenkwalder}},
  \bibinfo{author}{\bibfnamefont{C.}~\bibnamefont{Kohstall}},
  \bibinfo{author}{\bibfnamefont{M.}~\bibnamefont{Zaccanti}},
  \bibinfo{author}{\bibfnamefont{D.}~\bibnamefont{Naik}},
  \bibinfo{author}{\bibfnamefont{A.~I.} \bibnamefont{Sidorov}},
  \bibinfo{author}{\bibfnamefont{F.}~\bibnamefont{Schreck}}, \bibnamefont{and}
  \bibinfo{author}{\bibfnamefont{R.}~\bibnamefont{Grimm}},
  \bibinfo{journal}{Phys. Rev. Lett.} \textbf{\bibinfo{volume}{106}},
  \bibinfo{pages}{115304} (\bibinfo{year}{2011}).

\bibitem[{\citenamefont{Levinsen et~al.}(2009)\citenamefont{Levinsen, Tiecke,
  Walraven, and Petrov}}]{Levinsen2009ads}
\bibinfo{author}{\bibfnamefont{J.}~\bibnamefont{Levinsen}},
  \bibinfo{author}{\bibfnamefont{T.~G.} \bibnamefont{Tiecke}},
  \bibinfo{author}{\bibfnamefont{J.~T.~M.} \bibnamefont{Walraven}},
  \bibnamefont{and} \bibinfo{author}{\bibfnamefont{D.~S.}
  \bibnamefont{Petrov}}, \bibinfo{journal}{Phys. Rev. Lett.}
  \textbf{\bibinfo{volume}{103}}, \bibinfo{pages}{153202}
  (\bibinfo{year}{2009}).

\bibitem[{\citenamefont{Levinsen and Petrov}(2011)}]{Levinsen2011ada}
\bibinfo{author}{\bibfnamefont{J.}~\bibnamefont{Levinsen}} \bibnamefont{and}
  \bibinfo{author}{\bibfnamefont{D.}~\bibnamefont{Petrov}},
  \bibinfo{journal}{Eur. Phys. J. D} \textbf{\bibinfo{volume}{65}},
  \bibinfo{pages}{67} (\bibinfo{year}{2011}).

\bibitem[{\citenamefont{Kohstall et~al.}(2012)\citenamefont{Kohstall, Zaccanti,
  Jag, Trenkwalder, Massignan, Bruun, Schreck, and Grimm}}]{Kohstall2012mac}
\bibinfo{author}{\bibfnamefont{C.}~\bibnamefont{Kohstall}},
  \bibinfo{author}{\bibfnamefont{M.}~\bibnamefont{Zaccanti}},
  \bibinfo{author}{\bibfnamefont{M.}~\bibnamefont{Jag}},
  \bibinfo{author}{\bibfnamefont{A.}~\bibnamefont{Trenkwalder}},
  \bibinfo{author}{\bibfnamefont{P.}~\bibnamefont{Massignan}},
  \bibinfo{author}{\bibfnamefont{G.~M.} \bibnamefont{Bruun}},
  \bibinfo{author}{\bibfnamefont{F.}~\bibnamefont{Schreck}}, \bibnamefont{and}
  \bibinfo{author}{\bibfnamefont{R.}~\bibnamefont{Grimm}},
  \bibinfo{journal}{Nature} \textbf{\bibinfo{volume}{485}},
  \bibinfo{pages}{615} (\bibinfo{year}{2012}).

\bibitem[{\citenamefont{Pauling}(1928)}]{pauling1928tao}
\bibinfo{author}{\bibfnamefont{L.}~\bibnamefont{Pauling}},
  \bibinfo{journal}{Chem. Rev.} \textbf{\bibinfo{volume}{5}},
  \bibinfo{pages}{173} (\bibinfo{year}{1928}).

\bibitem[{\citenamefont{Sobelman}(1972)}]{sobelman1973ait}
\bibinfo{author}{\bibfnamefont{I.~I.} \bibnamefont{Sobelman}},
  \emph{\bibinfo{title}{An introduction to the theory of atomic spectra}}
  (\bibinfo{publisher}{Pergamon Press, Oxford}, \bibinfo{year}{1972}).

\bibitem[{\citenamefont{Baranger}(1958{\natexlab{a}})}]{baranger1958sqm}
\bibinfo{author}{\bibfnamefont{M.}~\bibnamefont{Baranger}},
  \bibinfo{journal}{Phys. Rev.} \textbf{\bibinfo{volume}{111}},
  \bibinfo{pages}{481} (\bibinfo{year}{1958}{\natexlab{a}}).

\bibitem[{\citenamefont{Baranger}(1958{\natexlab{b}})}]{baranger1958git}
\bibinfo{author}{\bibfnamefont{M.}~\bibnamefont{Baranger}},
  \bibinfo{journal}{Phys. Rev.} \textbf{\bibinfo{volume}{112}},
  \bibinfo{pages}{855} (\bibinfo{year}{1958}{\natexlab{b}}).

\bibitem[{MJS()}]{MJSupMat}
\bibinfo{note}{See Supplemental Material at [URL will be inserted by publisher]
  for details on the calibration of the FR center, the preparation of the
  atom-dimer mixture, the determination of the temperatures and the densities,
  and the effect of higher partial waves on the forward-scattering amplitude.}

\bibitem[{\citenamefont{Spiegelhalder et~al.}(2010)\citenamefont{Spiegelhalder,
  Trenkwalder, Naik, Kerner, Wille, Hendl, Schreck, and
  Grimm}}]{Spiegelhalder2010aop}
\bibinfo{author}{\bibfnamefont{F.~M.} \bibnamefont{Spiegelhalder}},
  \bibinfo{author}{\bibfnamefont{A.}~\bibnamefont{Trenkwalder}},
  \bibinfo{author}{\bibfnamefont{D.}~\bibnamefont{Naik}},
  \bibinfo{author}{\bibfnamefont{G.}~\bibnamefont{Kerner}},
  \bibinfo{author}{\bibfnamefont{E.}~\bibnamefont{Wille}},
  \bibinfo{author}{\bibfnamefont{G.}~\bibnamefont{Hendl}},
  \bibinfo{author}{\bibfnamefont{F.}~\bibnamefont{Schreck}}, \bibnamefont{and}
  \bibinfo{author}{\bibfnamefont{R.}~\bibnamefont{Grimm}},
  \bibinfo{journal}{Phys. Rev. A} \textbf{\bibinfo{volume}{81}},
  \bibinfo{pages}{043637} (\bibinfo{year}{2010}).

\bibitem[{\citenamefont{Naik et~al.}(2011)\citenamefont{Naik, Trenkwalder,
  Kohstall, Spiegelhalder, Zaccanti, Hendl, Schreck, Grimm, Hanna, and
  Julienne}}]{Naik2011fri}
\bibinfo{author}{\bibfnamefont{D.}~\bibnamefont{Naik}},
  \bibinfo{author}{\bibfnamefont{A.}~\bibnamefont{Trenkwalder}},
  \bibinfo{author}{\bibfnamefont{C.}~\bibnamefont{Kohstall}},
  \bibinfo{author}{\bibfnamefont{F.~M.} \bibnamefont{Spiegelhalder}},
  \bibinfo{author}{\bibfnamefont{M.}~\bibnamefont{Zaccanti}},
  \bibinfo{author}{\bibfnamefont{G.}~\bibnamefont{Hendl}},
  \bibinfo{author}{\bibfnamefont{F.}~\bibnamefont{Schreck}},
  \bibinfo{author}{\bibfnamefont{R.}~\bibnamefont{Grimm}},
  \bibinfo{author}{\bibfnamefont{T.}~\bibnamefont{Hanna}}, \bibnamefont{and}
  \bibinfo{author}{\bibfnamefont{P.}~\bibnamefont{Julienne}},
  \bibinfo{journal}{Eur. Phys. J. D} \textbf{\bibinfo{volume}{65}},
  \bibinfo{pages}{55} (\bibinfo{year}{2011}).

\bibitem[{\citenamefont{Petrov}(2004)}]{Petrov2004tbp}
\bibinfo{author}{\bibfnamefont{D.~S.} \bibnamefont{Petrov}},
  \bibinfo{journal}{Phys. Rev. Lett.} \textbf{\bibinfo{volume}{93}},
  \bibinfo{pages}{143201} (\bibinfo{year}{2004}).

\bibitem[{MJE({\natexlab{a}})}]{MJEndnote}
\bibinfo{note}{To determine the peak shift and the width, we apply a
  double-Gaussian fit to the spectra. From the fit, we identify the rf detuning
  of maximum signal and the width.}

\bibitem[{MJE({\natexlab{b}})}]{MJEndnote2}
\bibinfo{note}{The finite duration of our rf pulse causes an additional
  Gaussian broadening of typically $1.2\,$kHz (FWHM).}

\bibitem[{\citenamefont{Szudy and Baylis}(1996)}]{szudy1996pol}
\bibinfo{author}{\bibfnamefont{J.}~\bibnamefont{Szudy}} \bibnamefont{and}
  \bibinfo{author}{\bibfnamefont{W.~E.} \bibnamefont{Baylis}},
  \bibinfo{journal}{Phys. Rep.} \textbf{\bibinfo{volume}{266}},
  \bibinfo{pages}{127} (\bibinfo{year}{1996}).

\bibitem[{\citenamefont{Ferlaino et~al.}(2011)\citenamefont{Ferlaino, Zenesini,
  Berninger, Huang, N\"{a}gerl, and Grimm}}]{Ferlaino2011eri}
\bibinfo{author}{\bibfnamefont{F.}~\bibnamefont{Ferlaino}},
  \bibinfo{author}{\bibfnamefont{A.}~\bibnamefont{Zenesini}},
  \bibinfo{author}{\bibfnamefont{M.}~\bibnamefont{Berninger}},
  \bibinfo{author}{\bibfnamefont{B.}~\bibnamefont{Huang}},
  \bibinfo{author}{\bibfnamefont{H.-C.} \bibnamefont{N\"{a}gerl}},
  \bibnamefont{and} \bibinfo{author}{\bibfnamefont{R.}~\bibnamefont{Grimm}},
  \bibinfo{journal}{Few-Body Syst.} \textbf{\bibinfo{volume}{51}},
  \bibinfo{pages}{113} (\bibinfo{year}{2011}).

\bibitem[{\citenamefont{Nishida}(2009)}]{Nishida2009cia}
\bibinfo{author}{\bibfnamefont{Y.}~\bibnamefont{Nishida}},
  \bibinfo{journal}{Phys. Rev. A} \textbf{\bibinfo{volume}{79}},
  \bibinfo{pages}{013629} (\bibinfo{year}{2009}).
  
\bibitem[{\citenamefont{Chin and Julienne}(2005)}]{Chin2005rft}
\bibinfo{author}{\bibfnamefont{C.}~\bibnamefont{Chin}} \bibnamefont{and}
  \bibinfo{author}{\bibfnamefont{P.~S.} \bibnamefont{Julienne}},
  \bibinfo{journal}{Phys. Rev. A} \textbf{\bibinfo{volume}{71}},
  \bibinfo{eid}{012713} (\bibinfo{year}{2005}).

\end{thebibliography}

\clearpage

\end{document}